\documentclass{article}

\usepackage{amsmath}
\usepackage[utf8]{inputenc}
\usepackage{amssymb}
\usepackage{amsthm}
\usepackage{dsfont}
\usepackage{graphicx}
\usepackage{dsfont,amsthm,enumerate,amsfonts,caption,subcaption}
\usepackage{float}
\usepackage{hyperref}
\usepackage[affil-it]{authblk}

\newtheorem{corollary}{Corollary}

\title{The difference between two random mixed quantum states: exact and asymptotic spectral analysis }
\author[1]{José Mejía%
	\thanks{Electronic address: \texttt{jr.mejia1228@uniandes.edu.co}; José Mejía}}

\author[2]{Camilo Zapata%
	\thanks{Electronic address: \texttt{zcamilo@phys.ethz.ch}; Camilo Zapata}}

\author[1]{Alonso Botero%
	\thanks{Electronic address: \texttt{abotero@uniandes.edu.co}; Alonso Botero}}
\affil[1]{Departamento de F\'isica, Universidad de los Andes, Cra 1E No 18A-12, Bogot\'a,  Colombia}
\affil[2]{Department of Physics, ETH Zürich, Otto-Stern-Weg 1, 8093 Zürich, Switzerland}

\newcommand{\ket}[1]{\ensuremath{\left|#1\right\rangle}}

\DeclareMathOperator{\Tr}{Tr}
\DeclareMathOperator{\density}{\varrho}

\begin{document}

\maketitle

\begin{abstract}
We investigate the spectral statistics  of the  difference of  two  density matrices, each of which is independently obtained by partially tracing a random bipartite  pure quantum state.  We first show how a closed-form expression for the exact joint eigenvalue probability density function for arbitrary dimensions can be obtained from the joint probability density function of the diagonal elements of the difference matrix, which is straightforward to compute. Subsequently, we use standard results from free probability theory to  derive a relatively simple analytic expression for the asymptotic eigenvalue density (AED) of the difference matrix ensemble, and using  Carlson's theorem, we obtain an expression for its absolute moments. These results allow us to quantify the typical asymptotic distance between the two random mixed states using various distance measures; in particular, we obtain the almost sure  asymptotic behavior of the operator norm distance and the trace distance.
\end{abstract}

\section{Introduction}

Among the key insights in quantum information that follow from the statistical properties of  quantum states in high-dimensional Hilbert spaces is that the overwhelming majority of pure bipartite  states are very close to being maximally entangled. More precisely, consider  pure, normalized, quantum states $\ket{\psi} \in \mathcal{H}_M \otimes \mathcal{H}_N$, where $\mathcal{H}_M$ and $\mathcal{H}_N$ are Hilbert spaces of finite dimensions $M$ and $N$ respectively, and for a given $\ket{\psi}$, let  $\rho$ be  its reduced density matrix in  $\mathcal{H}_N$; that is, $\rho = \mathrm{Tr}_{\mathcal{H}_M} |\psi\rangle \langle \psi |$ where  $\mathrm{Tr}_{\mathcal{H}_M}$ denotes the partial trace  over the Hilbert space $\mathcal{H}_M$. 
A standard measure of the entanglement of $\ket{\psi}$   is the so-called  \emph{entanglement entropy}, defined as  the von Neumann entropy $S = -\mathrm{Tr}_{\mathcal{H}_N}(\rho \log \rho) $,  so that  $S=0$ when $|{\psi}\rangle$ is separable and  $S = \log( \min(N,M) )$ when  $|\psi \rangle$ is maximally entangled.
A well known result,  motivated by an initial estimate of Lubkin  \cite{lubkin_entropy_1978}, later conjectured by Page  \cite{page_average_1993}, and finally proved by Foong and Kanno  \cite{foong_proof_1994} (see also  \cite{sanchez-ruiz_simple_1995, sen_average_1996}), is that the  average entanglement entropy over all pure states $\ket{\psi}$, when $N \leq M$,  is given by
\begin{equation}
\langle S \rangle = \sum_{k=M+1}^{M N}\frac{1}{k} - \frac{N-1}{2 M} ,
\label{Pageformula}
\end{equation}
which for $N \gg 1$ gives $\langle S \rangle \simeq \log N$, the maximal entanglement entropy, up to corrections of order $N/M$ (here, and henceforth, random states $|\psi\rangle$ are understood to be uniformly sampled on the hypersphere $\langle \psi| \psi \rangle=1$; that is, according to the pushforward of the $U(M N)$ Haar measure under the action of $U(M N)$ on a fixed  normalized vector of $\mathcal{H}_M \otimes \mathcal{H}_N$).
Since then,  a better understanding of the ubiquity of near-maximal entanglement in the bipartite setting  \cite{hayden_aspects_2006, dahlsten_entanglement_2014} has
been a subject of considerable interest, fuelled in part by its relevance to several important applications in quantum information (see e.g., \cite{majumdar_extreme_2010}), such as random quantum circuits  \cite{oliveira_generic_2007,plato_random_2008},
 superdense coding  \cite{harrow_superdense_2004, abeyesinghe_optimal_2006},
random quantum channels  \cite{hayden_counterexamples_2008, collins_random_2011,hastings_superadditivity_2009,brandao_hastings_2010,fukuda_entanglement_2010}
equilibrium thermodynamics  \cite{popescu_entanglement_2006, goldstein_canonical_2006,reimann_typicality_2007} and thermalization \cite{gemmer_quantum_2001,linden_quantum_2009,linden_speed_2010,hutter_dependence_2013,goldstein_extremely_2015}, among others.

One aspect of the problem that has earned particular attention is the  statistics of the spectrum of the partial density matrix $\rho$, the non-zero part of which corresponds to the so-called \emph{Schmidt spectrum}, the squares of the coefficients appearing in the Schmidt (i.e., singular value) decomposition of the state $|\psi \rangle$.
When $|\psi\rangle$ is sampled uniformly, as defined previously, the resulting ensemble of random reduced density matrices $\rho$ is an ensemble that is also known in the Random Matrix Theory (RMT) literarure as the Fixed Trace Wishart-Laguerre  (FTWL) ensemble with Dyson index $\beta=2$ \cite{nechita_asymptotics_2007, forrester_log-gases_2010}. The  joint probability density function (PDF) for the unordered Schmidt spectrum induced by the  uniformly-sampled pure state ensemble was first obtained by Lloyd and Pagels  \cite{lloyd_complexity_1988} and is given by
\begin{equation}
\density(\lambda_1,\lambda_2,...,\lambda_N) \propto \prod_{i=1}^{N} \lambda_i^{M-N}\prod_{i<j}(\lambda_i-\lambda_j)^2 ,
\label{Sommdistfast}
\end{equation}
where it is understood that $N \leq M$, $\sum_i \lambda_i = 1$ and $\lambda_i \geq 0$ (henceforth we shall use the symbol $\density$ for spectral probability densities and the standard $\rho$ for  density matrices). This PDF contains all the necessary information to derive, at least in principle, the statistical properties of any function of the quantum state that is invariant under local unitary transformations. Indeed, a fair amount of progress has been made in obtaining exact results for several quantities of interest, such as  moments and correlations of traces of powers of $\rho$ and its entropy  \cite{sommers_statistical_2004}, the average  eigenvalue PDF  \cite{kubotani_exact_2008, adachi_random_2009}, the smallest eigenvalue PDF \cite{znidaric_entanglement_2007, majumdar_exact_2008, chen_smallest_2010, akemann_compact_2011} and the largest eigenvalue PDFs  \cite{vivo_largest_2011}. For general values of $M$ and $N$ these exact formulas tend to be rather complicated and are therefore of limited practical use. However, considerable simplifications emerge in the asymptotic limit  $M,N \rightarrow \infty$, with the ratio $N/M$ fixed, using well-known asymptotic techniques in RMT. In particular,  it follows that the asymptotic eigenvalue density (AED) for the rescaled eigenvalues $x = N \lambda$, satisfies a Mar\v{c}enko-Pastur law \cite{marcenko_distribution_1967,page_average_1993}(see also  \cite{nadal_statistical_2011}), as in  the unconstrained Wishart ensemble, with parameters given by the ratio of dimensions $c = N/M$ :
\begin{equation}\label{marcenkopastur}
\tilde{\density}(x) = \max\left(1-\dfrac{1}{c},0\right)\delta(x)+\frac{1}{2 \pi c  x} \sqrt{ (x-x_-)(x_+-x) }I_{[x_-,x_+]}(x)dx,   ,
\end{equation}
where $x_\pm =  c\left( \frac{1}{\sqrt{c}} \pm 1\right)^2$, and $I_{[x_-,x_+]}(x)$ is the indicator function in the interval $[x_-,x_+]$. Reference  \cite{nadal_statistical_2011} also provides useful asymptotic results for the distribution of the Renyi entanglement entropies (including the von Neumann entropy), and the distribution of the largest Schmidt eigenvalue.

Motivated by the general question of how random bipartite states concentrate around the maximally entangled state, in this paper we address a closely related problem: suppose two random bipartite states $|\psi_1 \rangle $ and $|\psi_2 \rangle$ are uniformly sampled independently from $\mathcal{H}_M \otimes \mathcal{H}_N$, and their corresponding partial density matrices   $\rho_1 = \mathrm{Tr}_{\mathcal{H}_M}\left(|\psi_1 \rangle \langle \psi_1 |\right)$ and $\rho_2 = \mathrm{Tr}_{\mathcal{H}_M}\left(|\psi_2 \rangle \langle \psi_2 |\right)$  are computed in $\mathcal{H}_N$. For appropriately large dimensions, with $N \leq M$, we expect $\rho_1 \simeq \rho_2 \simeq \mathds{1}/N$. Our question is then: how close are the states $\rho_1$ and $\rho_2$ from each other? More specifically, we will examine the eigenvalue statistics for the difference matrix
$
Z \equiv \rho_1 - \rho_2 \, ,
$
from which various  distance measures can be calculated.  In fact, the ensemble of matrices $Z$ defines a unitarily invariant random 
matrix ensemble, a fact that automatically implies that $\rho_1$ and  $\rho_2$ cannot be ``too close'', given the well-known phenomenon of eigenvalue repulsion that characterizes unitarily invariant ensembles, among others. Moreover, unitary invariance  allows us to use powerful tools to derive exact, though cumbersome, expressions for the joint PDF of the eigenvalues  for finite $N,M$,  as well as a relatively simple formula for the AED and its moments in the asymptotic limit  $N,M\to\infty$ with a fixed ratio $N/M$. Thus, the purpose of this paper is twofold: First, we derive  the exact expression for the joint eigenvalue PDF in the finite $N,M$ case, and in the asymptotic limit,  obtain closed form expressions for the AED and its moments; second, using these results, we derive the almost sure asymptotic  values of the trace distance $d_{\mathrm{tr}}(\rho_1,\rho_2) = \frac{1}{2} \sum_{\lambda \in \mathrm{spec}{Z}}| \lambda|$  and the operator norm distance $\|\rho_1-\rho_2\|_{\mathrm{op}}  = \sup_{\lambda \in spec(Z)} |\lambda|$, both of which are especially relevant for applications to quantum information theory.

The structure and summary of results of the paper is as follows: In Section \ref{section_results} we present  our main results in the form of three new theorems (Theorems \ref{exact_theorem}, \ref{disttheorem}, \ref{momentsth}) and two corollaries (Corollaries \ref{corollary_op_norm}, \ref{corollary}), and the remaining sections are devoted to the proofs of these results. The first result is  a closed-form formula for the joint eigenvalue PDF for the difference matrix ensemble, which follows from applying a powerful technique of Christandl \emph{et al.}\cite{christandl_eigenvalue_2014}, the so-called \emph{derivative principle}, which we reproduce here as Theorem \ref{derivativethm}. The derivative principle provides a connection between the  joint eigenvalue PDF and the joint PDF of the diagonal matrix elements of a unitarily invariant random matrix ensemble that  is then exploited in our first main result, Theorem \ref{exact_theorem}, to obtain a closed form expression for the joint eigenvalue PDF in terms  of  associated Laguerre polynomials, valid for arbitrary dimensions $N$ and $M$ with $N \leq M$. The resulting formula is quite complicated, but may nevertheless prove useful for small $N$ and $M$; in particular, the specialization to the case $N=2$ and general $M$ yields a relatively tractable formula. Section \ref{section_exact} provides a proof of these results, including an alternative proof of the derivative principle that uses standard Random Matrix Theory techniques. The next set of results are concerned with  the asymptotics of the eigenvalue PDF of  the difference matrix $Z$. Theorem \ref{disttheorem}  gives the AED for  $N, M \rightarrow \infty$ for all fixed values of the ratio $c = N/M$  (here, the constraint $N\leq M$ is relaxed), a result that is proved in Section \ref{section_Asymptotic} using standard results from Free Probability Theory \cite{voiculescu_free_1992}.  The AED  shows an interesting transition at the critical value $c=2$. For values of $c < 2$, the AED has positive support in the region $|x|<x_+$, with $x_+^2=\frac{1}{16}(\sqrt{4c+1}+3)^3(\sqrt{4c+1}-1)$, whereas for values of $c \geq 2$, the AED has positive support in the two regions defined by $x_-<|x|<x_+$, with $x_-^2=\frac{1}{16}(\sqrt{4c+1}-3)^3(\sqrt{4c+1}+1)$, and  a Dirac point measure at the origin.
Our other main result, presented in Theorem \ref{momentsth}, is an expression for the absolute moments (including moments of complex order) of the AED, again for all values of $c$, a result that is proved in Section \ref{section_Moments} using Carlson's theorem. 
The two corollaries to our main results involve the asymptotic almost sure behavior of two distance measures between the independent partial density matrices $\rho_1$ and $\rho_2$, namely the operator norm distance (Corollary \ref{corollary_op_norm}) , which is obtained from the upper support point $x_+$ of the AED, and the trace norm distance (Corollary \ref{corollary}), which follows from Theorem \ref{momentsth} when specialized to the first absolute moment.

\section{Main Results}\label{section_results}

Let $\mathcal{H}_M$ and $\mathcal{H}_N$ be two Hilbert spaces of dimensions $M$ and $N$ respectively.  Now let $\ket{\psi_1}$ and $\ket{\psi_2}$ be two normalized random pure states in the tensor product Hilbert space $\mathcal{H}_M \otimes \mathcal{H}_N$, which are uniformly and  independently sampled, and define $\rho_1$ and $\rho_2$ as the corresponding reduced density matrices for the system $\rho_1 = \mathrm{Tr}_{\mathcal{H}_M}( |\psi_1 \rangle \langle \psi_1 |)$ and $\rho_2 = \mathrm{Tr}_{\mathcal{H}_M}( |\psi_2 \rangle \langle \psi_2 |)$. Finally, define the difference matrix $Z$ as
 \begin{equation}
 Z  \equiv \rho_1 - \rho_2 \, .
 \end{equation}
For this difference matrix ensemble, our main results are concerned with the exact joint eigenvalue probability PDF $\density(\vec{\lambda})$, where $\vec{\lambda}=(\lambda_1, \lambda_2, \ldots \lambda_N)$ are the (unordered) eigenvalues of $Z$, and the AED, the single-eigenvalue marginal of $\density(\vec{\lambda})$ in the asymptotic limit $(N,M) \rightarrow \infty$, with fixed ratio $N/M$.

\subsection{The Exact Joint Eigenvalue Density}

We begin with the exact joint eigenvalue  PDF for all $N,M$, but with the provision that $N \leq M$. The result is presented in the form of two theorems, of which the first is a recasting of a previous result and the second one is  the original result. Both theorems will be proved in Section \ref{section_exact}.

Theorem \ref{derivativethm} establishes an extremely  useful connection  between the joint PDF of the eigenvalues of a unitarily-invariant random matrix ensemble and the joint PDF of the matrix diagonal elements of the same ensemble, which in general is considerably simpler to compute. This result is known in more general form from the  theory of Duistermaat-Heckman (DH) measures \cite{heckman_projections_1982,guillemin_moment_1994} which are measures on the dual Lie algebra of a  Lie group with a Hamiltonian action on a symplectic manifold, obtained from the push-forward of the Liouville measure along the corresponding moment map.  It concerns the connection between the so-called \emph{non-abelian} DH-measure for the Hamiltonian action of a compact connected Lie group  and the \emph{abelian} DH-measure for its maximal torus  \cite{heckman_projections_1982, guillemin_heckman_1990}.
Recently, Christandl \emph{et al}  \cite{christandl_eigenvalue_2014} have used this connection, under the name of the \emph{derivative pinciple}, for the computation of the eigenvalue PDFs of reduced density matrices of  multipartite-entangled states. As this derivative principle is surprisingly not as well known as it should be in the context of random matrices, we recast it here and  prove it in the following section using the standard language of Random Matrix Theory:

\newtheorem{thm}{Theorem}

\begin{thm}[\emph{Derivative Principle for Unitarily Invariant Random Matrix Ensembles}\cite{christandl_eigenvalue_2014}]\label{derivativethm}
Let $Z$ be a random matrix drawn from a unitarily invariant random matrix ensemble,  $\density_Z$ the joint eigenvalue PDF for $Z$ and $\Psi_Z$  the joint PDF of the diagonal elements of $Z$. Then
\begin{equation}\label{distribucion}
\density_Z(\vec{\lambda}) = \left( \prod\limits_{p=1}^{N}p! \right)^{-1} \Delta( \vec{\lambda})\Delta\left(-\partial_{\vec{\lambda}}\right)\Psi_Z(\vec{\lambda}).
\end{equation}
where  $\Delta(\vec{\lambda})=\prod\limits_{i<j}(\lambda_j-\lambda_i)$ is the Vandermonde determinant and $\Delta(-\partial_{\vec{\lambda}}) $ the differential operator  $\prod_{i < j} \left(\frac{\partial}{\partial \lambda_i} -\frac{\partial}{\partial \lambda_j} \right)$.
\end{thm}

This theorem proves to be particularly useful to compute the joint eigenvalue PDF of the sum of independent random matrices drawn from unitarily invariant ensembles, given that the joint PDF of the diagonal elements is the convolution of the respective joint PDFs. This is precisely the case at hand for the ensemble of random matrices $Z = \rho_1 - \rho_2$, where $\rho_1$ and $\rho_2$ are the partial density matrices of two independent random bipartite states.

Theorem \ref{exact_theorem} gives a contour integral expression (or what equivalently can be cast as a constant term identity) for the joint PDF of the diagonal elements for our difference ensemble (an alternative expression in terms of Lauricella generalized hypergeometric functions is given at the end of Section \ref{Psi_subs}):

\begin{thm}\label{exact_theorem} Let  $Z = \rho_1 - \rho_2$ be the ensemble of difference matrices as defined at the beginning of the section. Then, the  joint PDF of diagonal elements $\vec{z} = (z_1,z_2,\ldots,z_N) \equiv (Z_{11},Z_{22},\ldots, Z_{NN})$ of $Z$  is given by the contour integral
\begin{equation}
\label{diagdist1}
\Psi_Z(\vec{z}) = \delta_R(\vec{z})\Gamma(MN)^2  \frac{(-1)^{N(M-1)}}{2 \pi i} \oint \! d s\frac{e^{s} }{s^{N(2M-1)}} \prod_{i=1}^{N} e^{-\frac{1}{2}|z_i|s} L_{M-1}^{1-2M
}(s |z_i|),
\end{equation}
where the contour encircles the origin, $L_n^{a}(x)$ are the associated Laguerre Polynomials,  $\delta_R(\vec{z})$  the surface delta function 
\begin{equation}\label{surface_delta}
\delta_R(\vec{z})=\delta\left(\sum_{i=1}^{N}z_i\right)I_R(\vec{z}),
\end{equation}
and $I_R(\vec{z})$ is the indicator function on the  $N-1$ dimensional region $R$ in $\mathbb{R}^N$ defined by the Minkowski difference set $R = S - S$, where $S$ is the standard probability simplex. 
\end{thm}

\begin{figure}[H]
\centering
\includegraphics[width=3.5in]{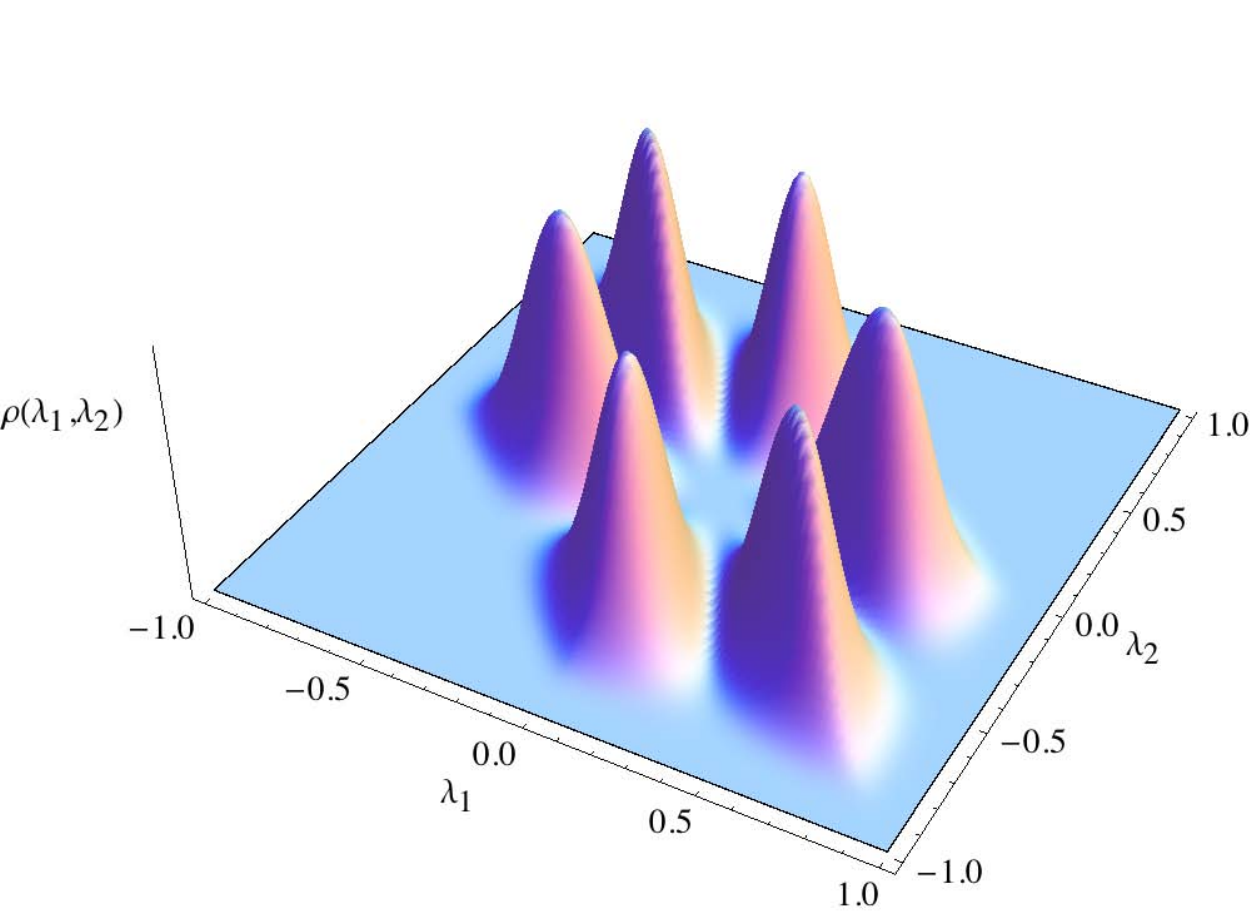}
  \caption{Joint PDF of eigenvalues $\density(\vec{\lambda})$ of $Z$ for the case $N=3$, $M=3$, as a function of two independent eigenvalues $\lambda_1$ and $\lambda_2$ (hence $\lambda_3 = - \lambda_1-\lambda_2$).  }
          \label{d3n3}
\end{figure}

Theorems \ref{derivativethm} and \ref{exact_theorem} provide a systematic, though  not necessarily practical, way of computing the joint eigenvalue PDF for the difference of two random partial density matrices, via equations \eqref{distribucion} and \eqref{diagdist1}.
In Fig. \ref{d3n3} we show the resulting PDF for the case $N=3$ and $M=3$.  The PDFs are supported within a convex polytope in the $\sum_{i=0}^{N} \lambda_i = 0$ hyperplane of $\mathbb{R}^N$, with vertices at $\vec{e}_i - \vec{e}_j$, $i \neq j$, where the $\vec{e}_i$ are the standard unit vectors in $\mathbb{R}^N$. They are symmetric under reflection ($\vec{\lambda} \rightarrow - \vec{\lambda}$) and permutation of the eigenvalues.  The multi-lobe shape of the PDF (see Fig. \ref{d3n3}) is a signature of the  well-known eigenvalue repulsion phenomenon, and is a consequence of  the Vandermonde determinant in \eqref{distribucion},  which forces the PDF to vanish whenever two of the eigenvalues are the same.

\begin{figure}[H]
\centering
        \begin{subfigure}{0.4 \textwidth}
                \includegraphics[width=\textwidth]{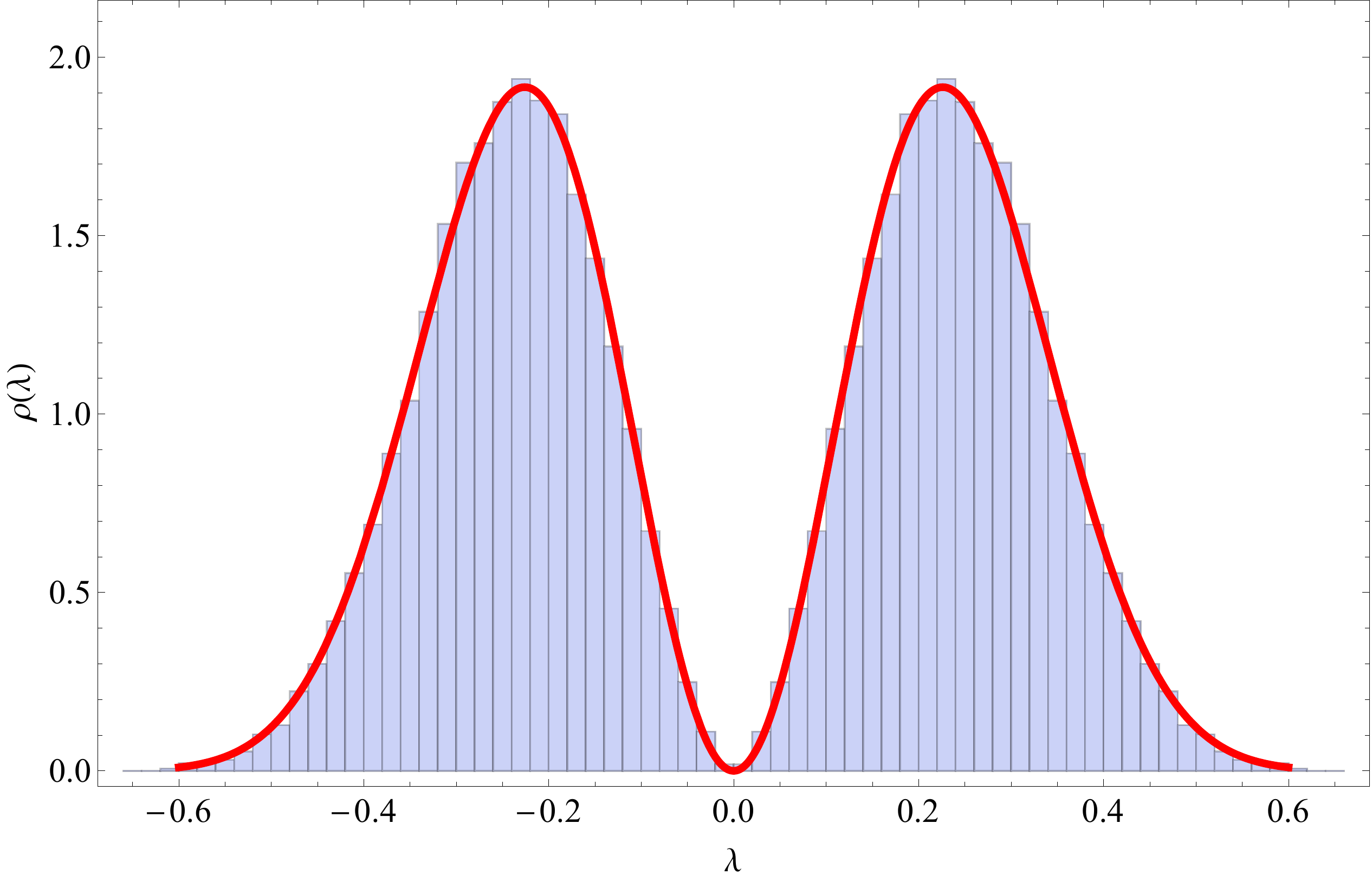}
                \caption{$N=2$ and $M=10$.}
                \label{n2m10}
        \end{subfigure}
        \quad
         \begin{subfigure}{0.4\textwidth}
                        \includegraphics[width=\textwidth]{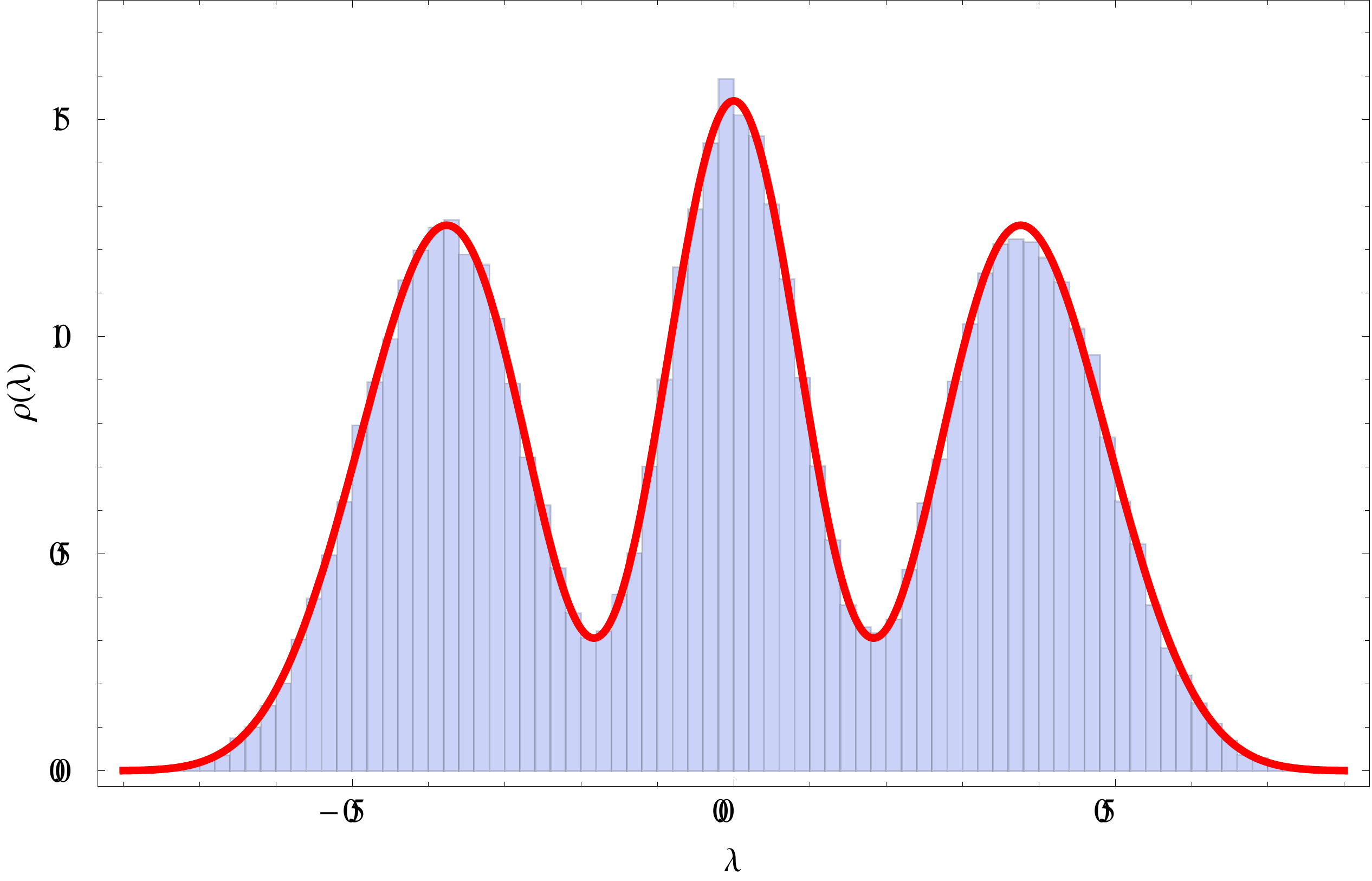}
                        \caption{$N=3$ and $M=3$.}
                        \label{n3m3}
          \end{subfigure}
     \\
          	\begin{subfigure}{0.4 \textwidth}
          		\includegraphics[width=\textwidth]{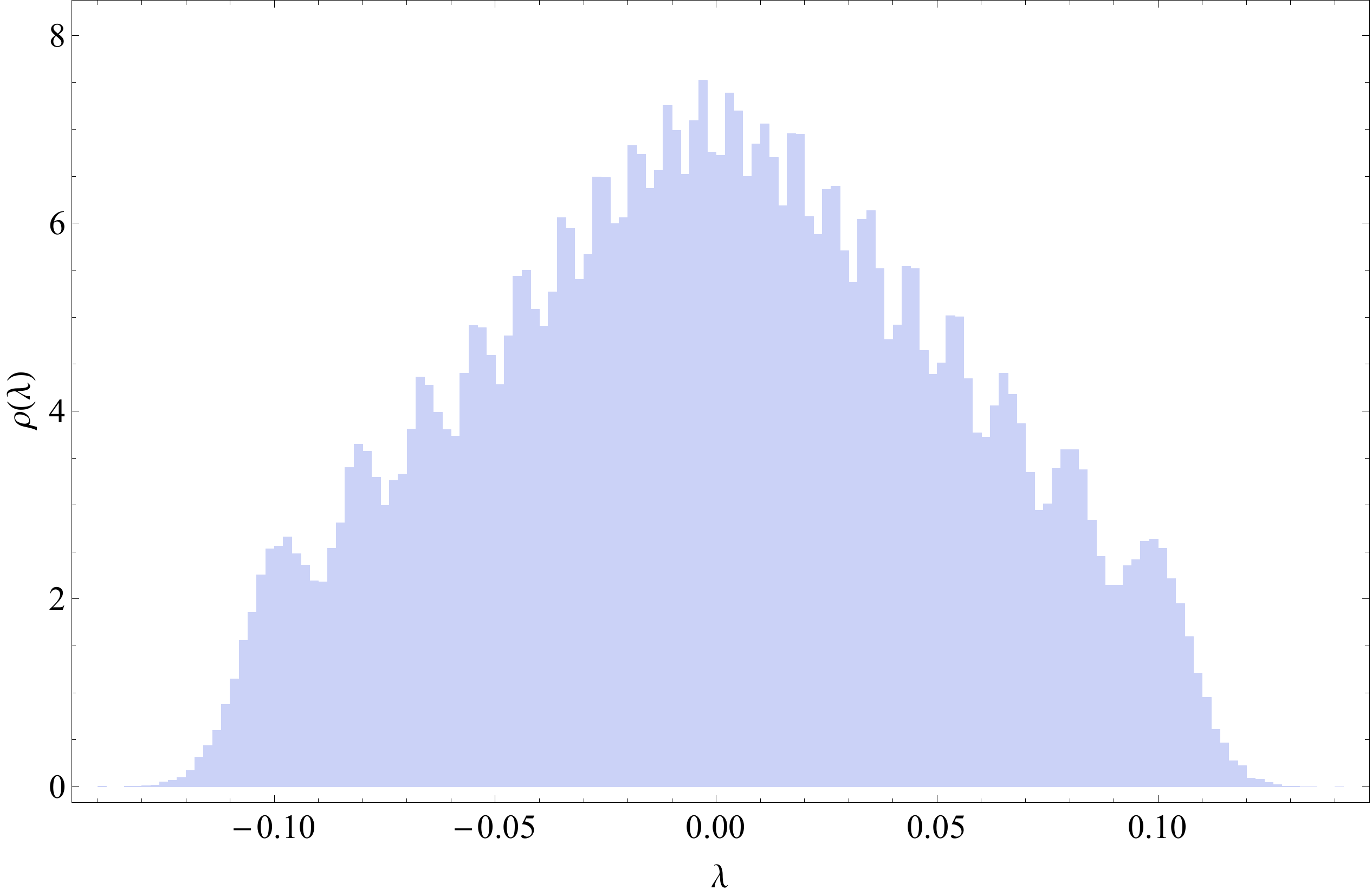}
          		\caption{$N=20$ and $M=35$.}
          		\label{n20m35}
          	\end{subfigure}
          	\quad
          	\begin{subfigure}{0.4\textwidth}
          		\includegraphics[width=\textwidth]{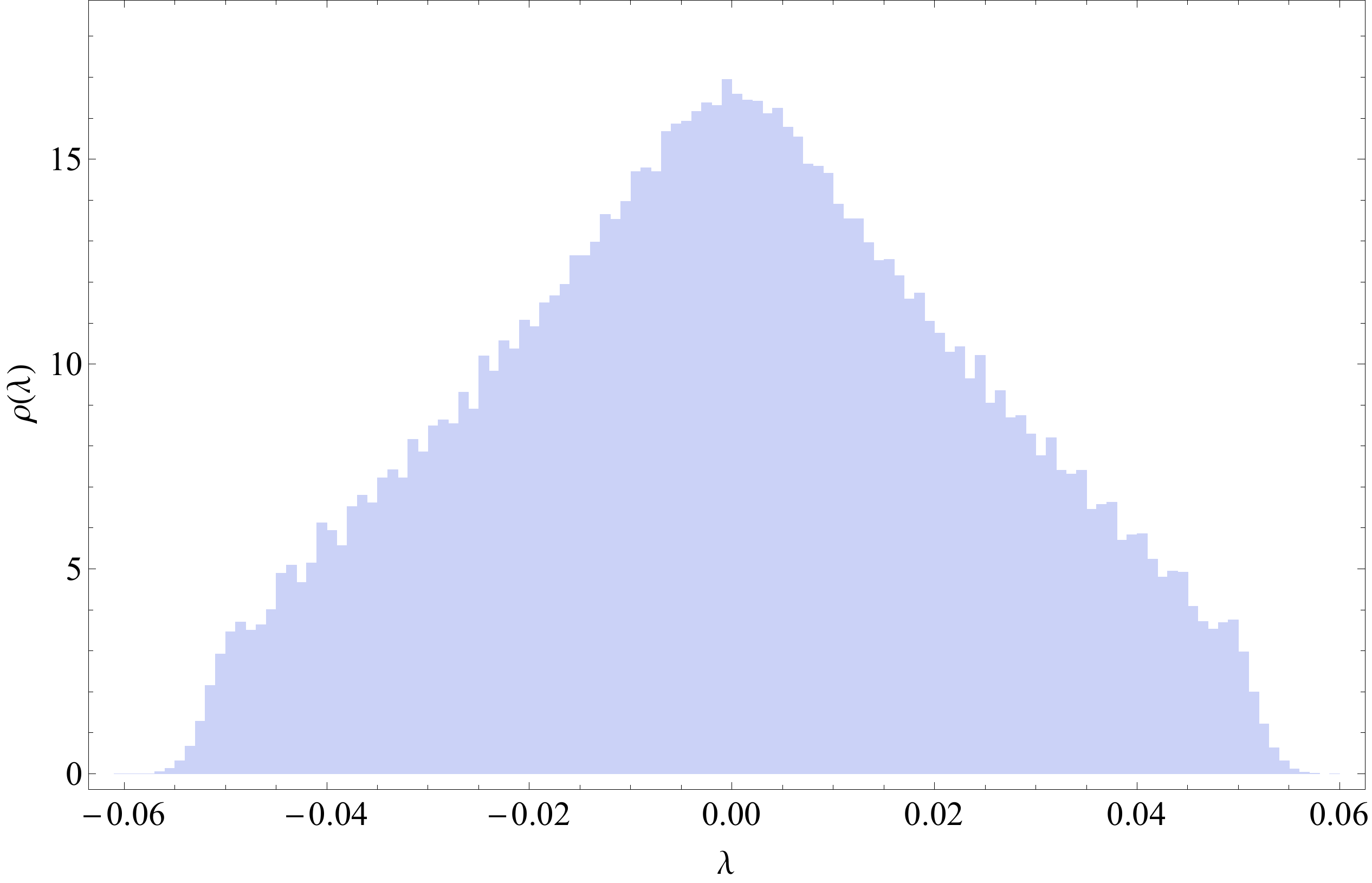}
          		\caption{$N=50$ and $M=70$.}
          		\label{n50m70}
          	\end{subfigure}
          	     \caption{Average eigenvalue density of $\rho_1-\rho_2$ for low dimensions (red line). The blue bars represent the normalized histograms obtained numerically from $30000$ random samples for Figs. \ref{n2m10} and \ref{n3m3}. For Figs. \ref{n20m35} and \ref{n50m70} the number of samples was $5000$.}
\label{figure_marginals1}
\end{figure}

Armed with the joint eigenvalue PDF, the average density of eigenvalues can be obtained by integrating out over $N-1$ eigenvalues. Unfortunately, there does not appear to be any particularly simple expressions for these marginal PDFs, except in the case $N=2$, where
 the exact result for the average eigenvalue PDF for any $M$ can be expressed in terms of a Gauss hypergeometric function ${}_2 F_1$ (see Appendix B) as:
\begin{equation}\label{d2equation}
\density(\lambda)= K_{M}\lambda \frac{\partial}{\partial \lambda}\left[ \frac{( |\lambda|^2-1 )^{2 M\!-\!1}}{ 1+|\lambda|} \, _2F_1\left(\begin{array}{c} 1/2 \ \ \ 1\!-\!M \\ 2(1\!-\! M) \end{array}\left | \frac{4 |\lambda|}{(1 + |\lambda|)^2}\right.\right)\right],
\end{equation}
where the proportionality constant is
\begin{equation}
K_{M}=\frac{\Gamma(2 M)^2 \Gamma( 2 M -1)^2 }{\Gamma(4M-2)\Gamma(M)^4}.
\end{equation}
In Figs. \ref{n2m10} and \ref{n3m3}  we show how our  results for $N=2, M=10$ and $N=3, M=3$ agree with empirical distributions obtained from independent samplings of $Z$.
From the empirical distributions shown in  Figs. \ref{n20m35} and \ref{n50m70} for higher values of $N$ (with $M \geq N$), it becomes evident that the average eigenvalue density will show  exactly $N$ peaks, which as $N$ grows become progressively less pronounced, tending to a smooth single-peaked, finitely-supported, distribution in the asymptotic limit $N \rightarrow \infty$.

\subsection{ The asymptotic eigenvalue density (AED)}

Considerable simplifications ensue in the asymptotic limit $(M,N) \rightarrow \infty$, with a fixed ratio $c =N/M$, if we concentrate on the   single-eigenvalue marginal of the joint eigenvalue PDF, to which the empirical eigenvalue density almost surely converges asymptotically. As we show in Section \ref{section_Asymptotic}, unitary invariance of the independent matrices $\rho_1$ and $\rho_2$ implies that  two ensembles satisfy the so-called freeness condition asymptotically \cite{voiculescu_free_1992}, from which it follows that the AED of $Z = \rho_1 - \rho_2$ can be obtained from the free convolution of the AEDs of $\rho_1$ and $-\rho_2$, which are given for all values of $c$ in terms of the Mar\v{c}enko-Pastur law, shown in equation \eqref{marcenkopastur}.  Our main result involves the asymptotic density function of the rescaled eigenvalues  $ x = N \lambda$, which we denote by  $\tilde{\density}(x)$:

 \begin{thm}
 \label{disttheorem}
 Let  $Z = \rho_1 - \rho_2$ be the ensemble of difference matrices as defined at the beginning of the section, and define the constants
 \begin{equation}\label{xpm}
 x_{\pm} = \frac{1}{4} \left(\sqrt{4 c + 1} \pm 3
 \right)^{3/2} \left( \sqrt{4 c + 1} \mp 1 \right )^{1/2} ,
 \end{equation}
 where $x_+$ is real and positive and $x_-$ purely imaginary for $c < 2$ and real and positive for $c > 2$. Then, in the limit $N \to \infty$, $M \to \infty$ with the ratio $c= N/M$ fixed, the  asymptotic rescaled eigenvalue density is, almost surely,
 \begin{equation}\label{roasymptotic1}
 \tilde{\density}(x)=\left \{ \begin{array}{lcl}
 \frac{\sqrt{(2-c)^2+3x^2}}{\sqrt{3}\pi c|x|}\sinh\left(\frac{l(x)}{3}\right) & {} & |x| < x_+ \\
 { } & { } & { } \\
 0 & { } & |x| \geq x_+
 \end{array}
 \right. ,
 \end{equation}
 for $c\leq 2$, and
 \begin{equation}\label{roasymptotic2}
 \tilde{\density}(x)=\left \{ \begin{array}{lcl}
 \frac{\sqrt{(2-c)^2+3x^2}}{\sqrt{3}\pi c|x|}\sinh\left(\frac{l(x)}{3}\right) & {} & |x| \in (x_-, x_+) \\
 { } & { } & { } \\
 \frac{c-2}{c} \delta(x)  & { } & |x| \notin (x_-, x_+)
 \end{array}
 \right. ,
 \end{equation}
 for $c>2$, where
 \begin{equation}
 l(x) = \log\left(\eta(x)+\sqrt{\eta^2(x)-1}\right),
 \end{equation}
 and
 \begin{equation}
  \eta(x) = \frac{9(c+1)x^2+(2-c)^3}{\left((2-c)^2+3x^2\right)^{3/2}}.
  \end{equation}
 \end{thm}

 Note that in expressions \eqref{roasymptotic1} and \eqref{roasymptotic2} we can alternatively write
  for the regions of positive support\begin{equation}\label{alternanive}
  \frac{\sqrt{(2-c)^2+3x^2}}{\sqrt{3}\pi c|x|}\sinh\left(\frac{l(x)}{3}\right) = \frac{1}{2 \pi c |x|}\left(w[x,c] - \frac{x^2+\frac{1}{3}(2-c)^2}{w[x,c]}\right),
  \end{equation}
  where
  \begin{equation}
  w(x)=
  \left(\sqrt{(x^2(x^2 - x_-^2)(x_+^2-x^2))}+ \sqrt{3}(c+1)\left(x^2+ \frac{(2-c)^3}{9 (c+1)}\right)\right)^{\frac{1}{3}}.
  \end{equation}

\begin{figure}[H]
  	\centering
  	\begin{subfigure}{0.4 \textwidth}
  		\includegraphics[width=\textwidth]{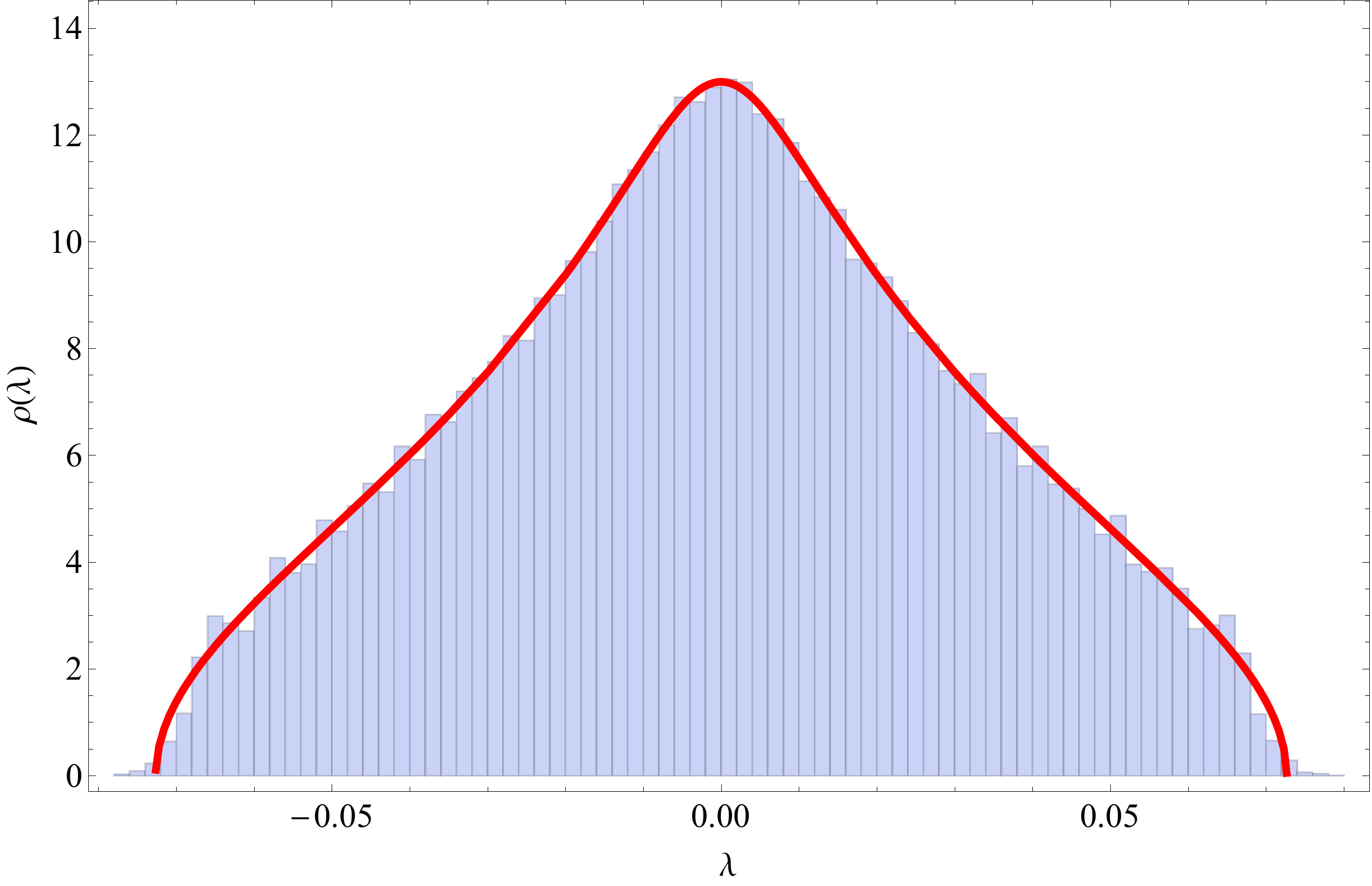}
  		\caption{$N=40$, $M=50$.}
  		\label{n35m20}
  	\end{subfigure}
  	\quad
  	\begin{subfigure}{0.4\textwidth}
  		\includegraphics[width=\textwidth]{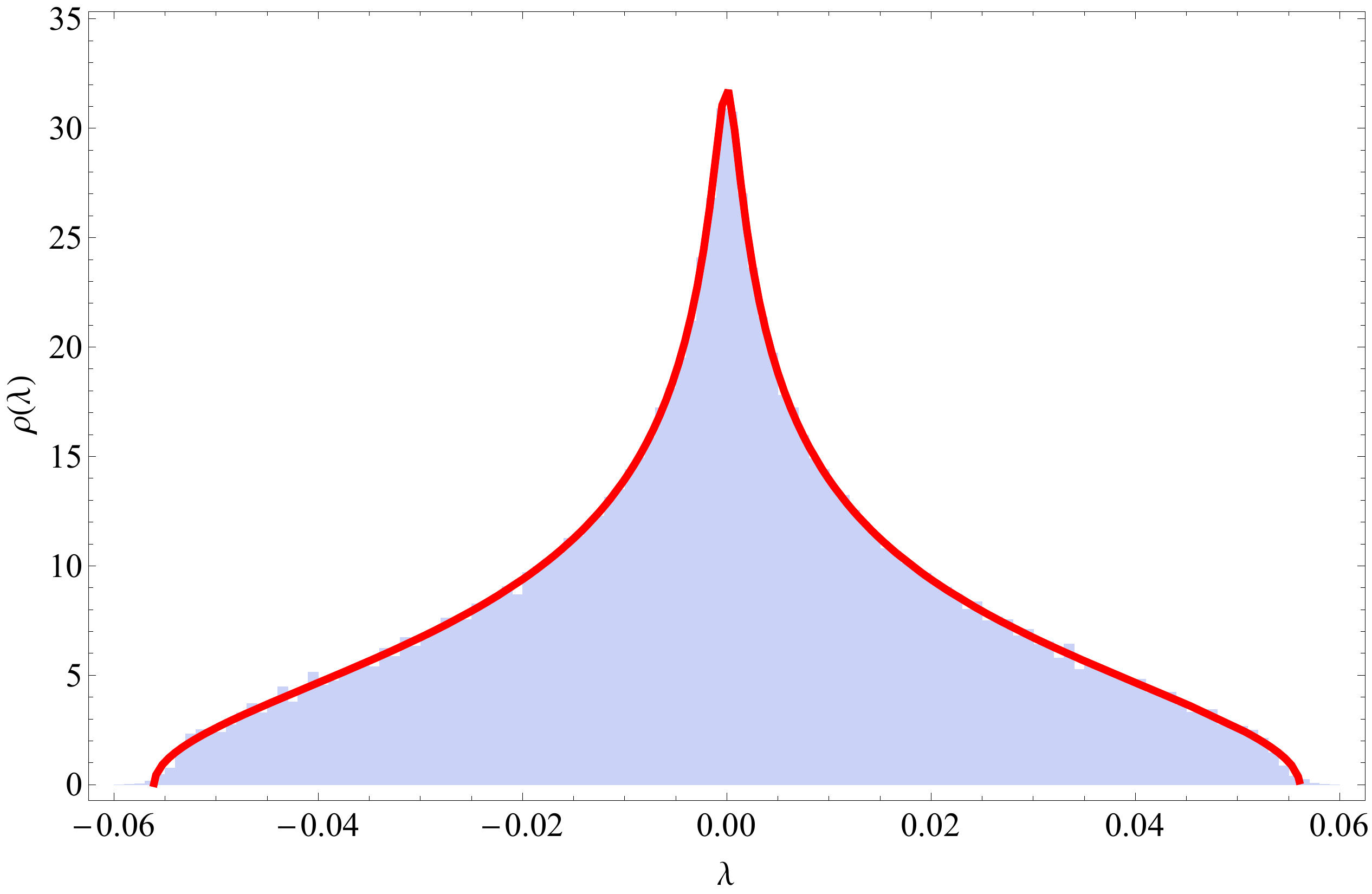}
  		\caption{$N=80$, $M=50$.}
  		\label{N80M50}
  	\end{subfigure}
  	\\
  	\begin{subfigure}{0.4 \textwidth}
  		\includegraphics[width=\textwidth]{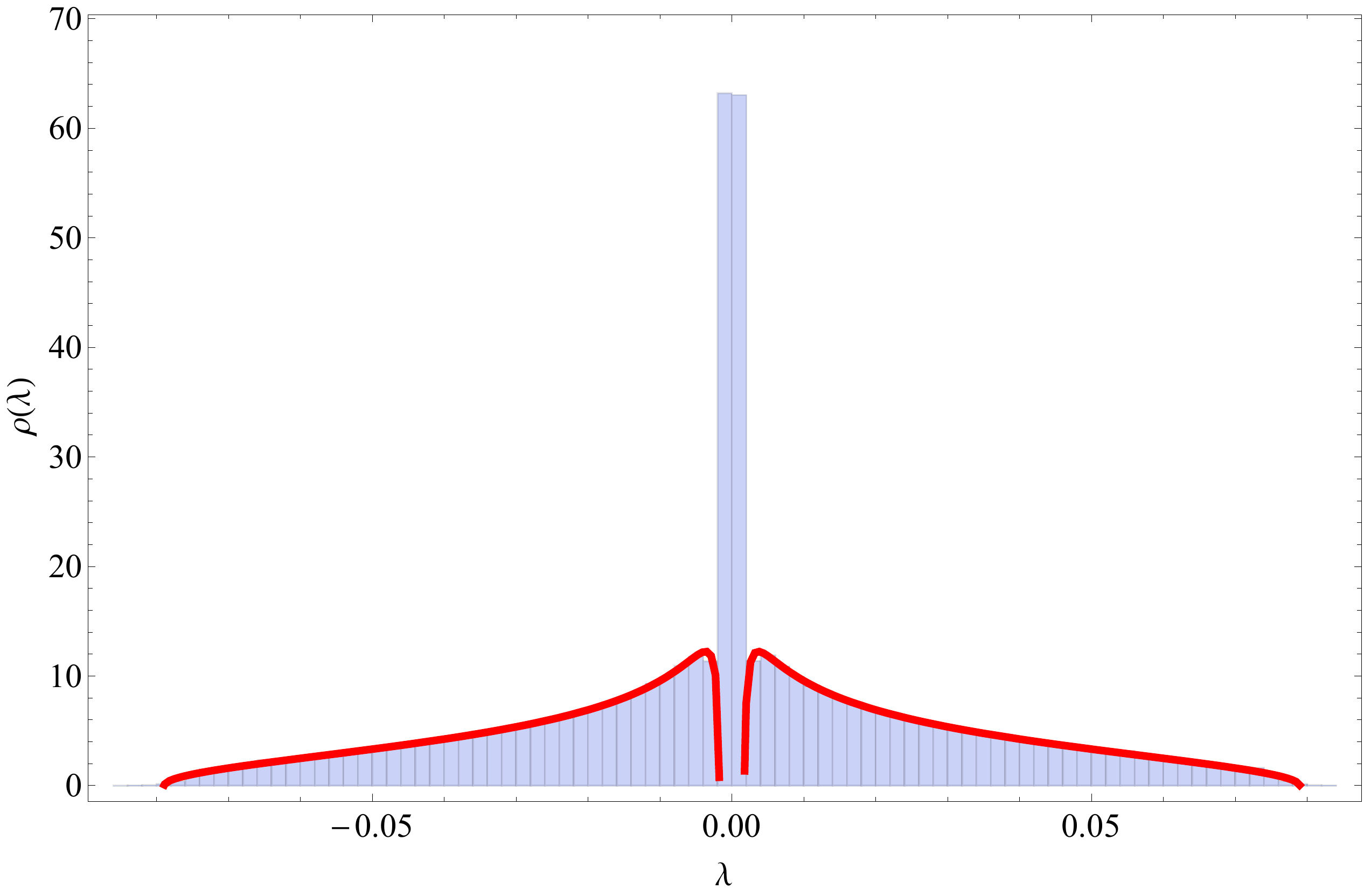}
  		\caption{$N=80$, $M=30$.}
  		\label{N80M30a}
  	\end{subfigure}
  	\quad
  	\begin{subfigure}{0.4\textwidth}
  		\includegraphics[width=\textwidth]{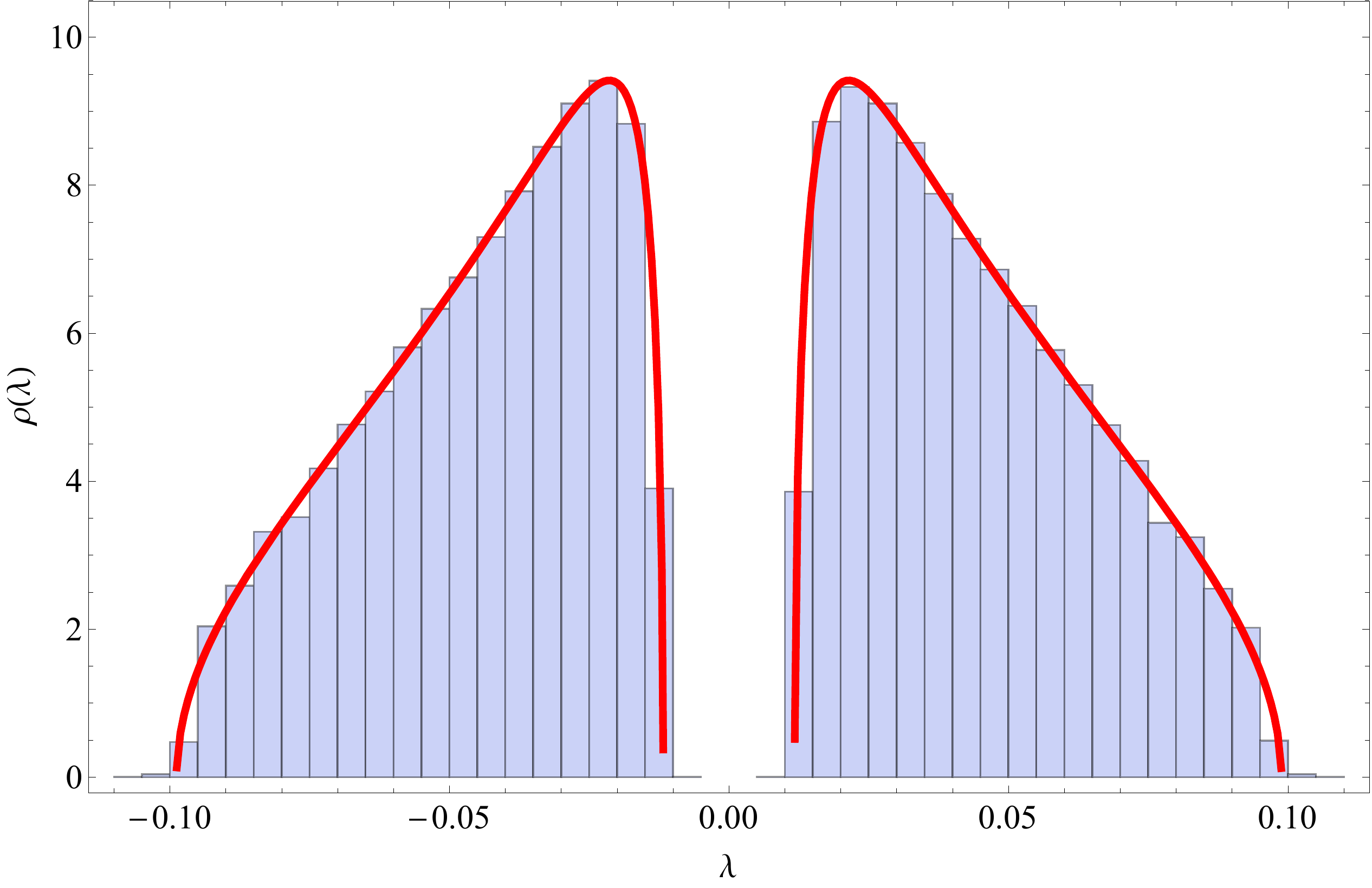}
  		\caption{$N=100$, $M=20$.}
  		\label{N100M20}
  	\end{subfigure}
  	\caption{Average eigenvalue density of $\rho_1-\rho_2$ for high dimensions (red line), compared to normalized histograms obtained numerically from $3000$ random samples. In Fig. \ref{N100M20} the values at 0 are not shown but constitute a fraction equal to $1-2/c=3/5$ of the total eigenvalues, as predicted. }
  	\label{figthmas}
  \end{figure}

 Fig. \ref{figthmas} shows how the AED depends on the parameter $c$. The most striking feature of this behavior is the fact that for $c>2$ a gap in the eigenvalue density arises for $|x|<x_-$, with a point distribution appearing at $x=0$ with weight $\frac{c-2}{c}$. The appearance of the point distribution for $c>2$ can be understood from the fact that if $2M<N$, the difference $\rho_1-\rho_2$ is of rank at most $2M$, and generically we expect the ranges of  $\rho_1$ and $\rho_2$ to be linearly  independent when $2M < N$, in which case the fraction of zero eigenvalues of $\rho_1-\rho_2$ should be $\frac{N-2M}{N}=1-\frac{2}{c}$. Also, from the asymptoic behavior of random subspaces (see Collins  \cite{collins_product_2005}), we expect that when $M/N \rightarrow 0$ ($c \rightarrow \infty$), the ranges of $\rho_1$ and $\rho_2$ should become asymptotically orthogonal and thus the non-zero eigenvalues of $\rho_1-\rho_2$ to be approximately the union of the non-zero eigenvalues of $\rho_1$ and $-\rho_2$. The AED of $\rho_1-\rho_2$ should then be the mixture of the AEDs of $\rho_1$ and $-\rho_2$, which follow the Mar\v{c}enko-Pastur law (or a reflected version of it).   Thus, the existence of a gap for $c>2$  reflects the gap that already exists in the Mar\v{c}enko-Pastur law in a region in the neighborhood of the origin, and the closing of the gap for $c<2$ may be understood as a consequence of strong mixing due to the fact that when $2M>N$ the ranges of $\rho_1$ and $\rho_2$ cannot be linearly independent subspaces. 
 
 An obvious variant of our problem is to consider differences  of the density matrices with different weights, \emph{i.e.}, $Z=p\rho_1-q\rho_2$.  The extension of the results of  Theorem \ref{disttheorem} for this more general class of ensembles can also be obtained in closed form, as is shown in Appendix A.

 \subsection{ Absolute Moments of the AED}
 
Our final theorem concerns the absolute moments of the  AED $\tilde{\density}(x)$ (Eqs. \eqref{roasymptotic1} and \eqref{roasymptotic2}).  As shown in  Section \ref{section_Moments}, the general theorem for the absolute moments  follows from Carlson's theorem, which warrants  an analytic extension of the closed-form  expression for the even integer moments that is obtained from the Laurent series of the Cauchy transform of  $\tilde{\density}(x)$. Our main result gives the absolute complex moments (or equivalently the Mellin transform for the density of $|x|$):
\begin{thm}\label{momentsth}
	Let $\tilde{\density}(x)$ be the AED  in Theorem \ref{disttheorem} and $z \in \mathbb{C}$ with $\text{Re}(z)>0$. Then the complex absolute moments of $\tilde{\density}(x)$,
	 	\begin{equation}\label{distance}
	m_z=\int_{-\infty}^{\infty}dx\ |x|^z \tilde{\density}(x),
	\end{equation}
	are
	\begin{equation}
m_{z}(c)=\left\{ \begin{array}{ll}
\frac{\Gamma(z+1) (2 c)^{z/2} }{\Gamma\left(\frac{z}{2}+1\right) \Gamma\left(\frac{z}{2}+2\right)}\, _2F_1\left(1-\frac{z}{2},-\frac{z}{2};\frac{z}{2}+2;\frac{c}{2}\right), & \ \ \ c\leq 2, \\
{} & {} \\
2 c^{z-1} \, _2F_1\left(1-\frac{z}{2},-z ;2;\frac{2}{c}\right) & \ \ \ c>2,
\end{array}
\right.
\label{formulas}
\end{equation}
where $_2F_1$ is the Gauss hypergeometric function.
\end{thm}

\subsection{Corollaries: Asymptotics of the  distance measures}

The main application of our results is in quantifying the distance between the two random states $\rho_1$ and $\rho_2$ using distance measures derived from the spectrum of $Z= \rho_1 - \rho_2$.  In particular, the almost sure behavior of two distance measures,  can be readily obtained as corollaries of Theorems \ref{disttheorem} and \ref{momentsth}. 

From the almost sure convergence of the empirical eigenvalue distributions to the AED  \cite{alsmeyer_asymptotic_2013}, the maximum absolute value of the eigenvalues of $Z$ almost surely converges to the $ x_+/N $, where $x_+$ is the upper support point of the AED. Hence, as a corollary to  Theorem \ref{disttheorem} we have:

\begin{corollary}\label{corollary_op_norm}
Under the conditions of Theorem \ref{disttheorem}, the operator norm of the difference $\rho_1 - \rho_2$,  almost surely behaves  as 
\begin{equation}
\| \rho_1 - \rho_2 \|_{op} \stackrel{a.s.}{\rightarrow} \frac{1}{4N}\left( \sqrt{1 +4 c} + 3  \right)^{3/2} \left( \sqrt{1 +4 c} -  1 \right)^{1/2}  .
\end{equation}
\end{corollary}
\noindent Note that for small $c$, the limiting value simplifies to $2 \sqrt{2 c}/N $. This may be compared to the leading behavior of  $\| \rho_1 -  \mathbb{I}/N\|_{op}$,  the  operator norm of the difference  between (say) the  random state $\rho_1$ and  its average, the totally mixed state $ \mathbb{I}/N$. Using the upper support point $x_+$ of the Mar\v{c}enko-Pastur law \eqref{marcenkopastur}, this leading behavior can be shown to be $\simeq 2 \sqrt{c}/N = 2 /\sqrt{N M}$, so that asymptotically $\| \rho_1 - \rho_2 \|_{op} \simeq \sqrt{2}\, \| \rho_1 -  \mathbb{I}/N \|_{op}$ for $c \ll 1$. Similarly, for $c \gg 1$, we find that both   $\| \rho_1 - \rho_2\|_{op}$ and $\| \rho_1 -\mathbb{I}/N \|_{op}$ behave as $\simeq c/N = 1/M$.

Next, we turn to the trace distance, which is defined as
\begin{equation}
d_{\mathrm{tr}}(\rho_1,\rho_2)=\dfrac{1}{2}\|\rho_1-\rho_2\|_1=\dfrac{1}{2}\Tr\sqrt{(\rho_1-\rho_2)(\rho_1-\rho_2)^\dagger},
\end{equation}
where $\|\cdot\|_1$ is the trace norm.  A straightforward consequence of the almost sure convergence of the AED (Theorem \ref{disttheorem}) is therefore that in the limit $N,M \rightarrow \infty$ (with $c$ constant), the trace distance tends to a limiting value
\begin{equation}
 d_{\mathrm{tr}}(\rho_1,\rho_2) \stackrel{a.s.}{\rightarrow} \dfrac{1}{2}\int_{-\infty}^{\infty}dx\ |x|\, \tilde{\density}(x).
\end{equation}
 Thus, the limiting trace distance follows as a corollary of Theorem \ref{momentsth} for  $z=1$,  together with hypergeometric identities (see Subsection \ref{absmom1}):
\begin{corollary}\label{corollary}
Under the conditions of Theorem \ref{disttheorem}, the trace distance between $\rho_1$ and $\rho_2$,  almost surely behaves  as 

\begin{equation}
d_{\mathrm{tr}}(\rho_1,\rho_2)\stackrel{a.s.}{\rightarrow}\left\{
\begin{array}{ll}
\dfrac{1}{2\pi c}\left((c+1) \sqrt{(2-c) c}+(4 c-2)\arcsin\left(\sqrt{\frac{c}{2}}\right)\right),& \ c \leq 2,\, \label{distance1}\\
{}& {}\\
 1-\frac{1}{2c},  &\ c>2.
\end{array}
\right.
\end{equation}
	
\end{corollary}

\begin{figure}[H]
	\centering
	\includegraphics[scale=0.4]{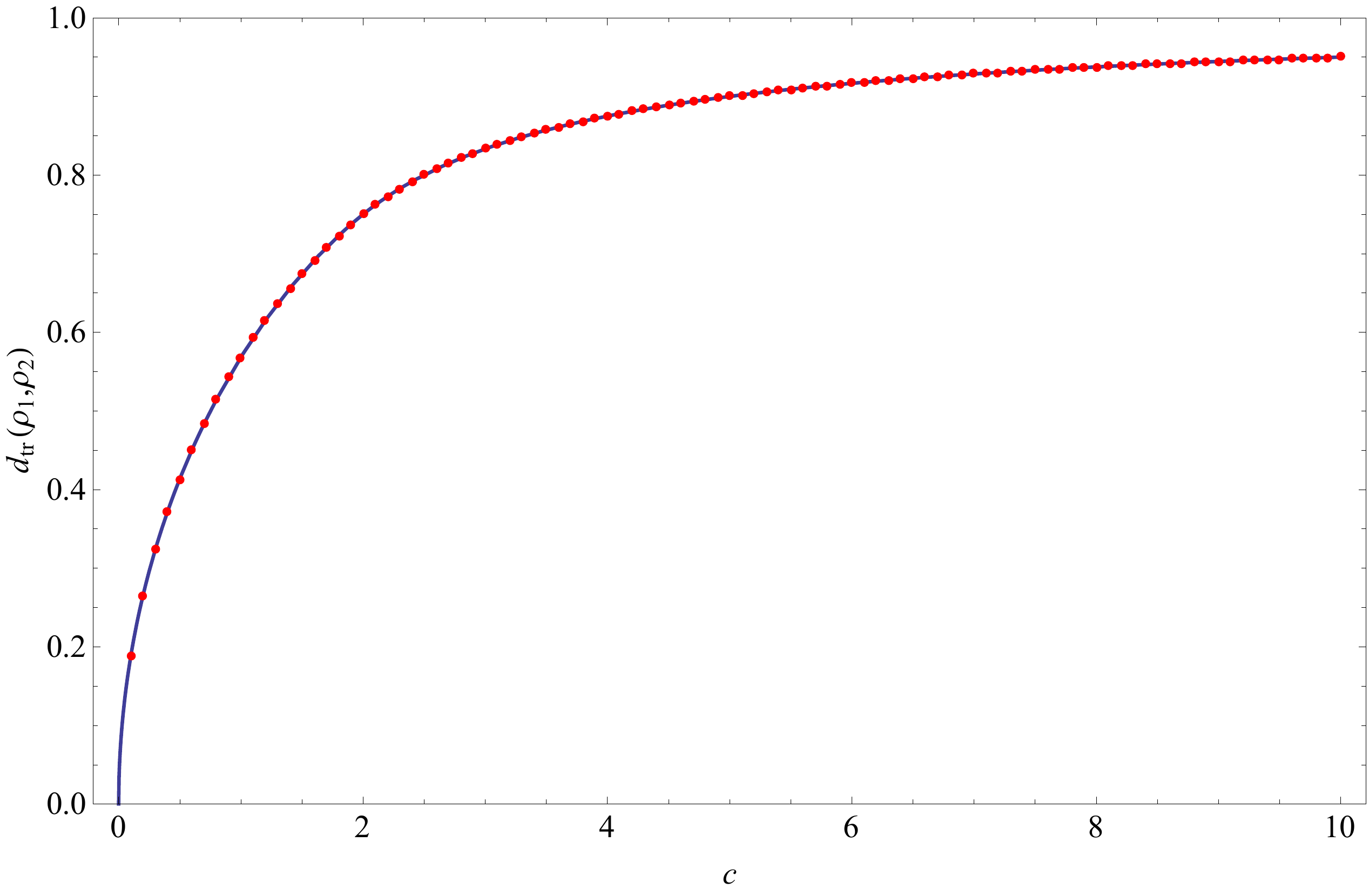}
	\caption{Trace distance $d_{tr}(\rho_1,\rho_2)$ as a function of $c$ (blue line) vs numerical values (red dots).}
	\label{pnormseven}
\end{figure}
As shown   in Fig. \ref{pnormseven}, the trace distance between two random states  asymptotically tends to the maximum value of $1$ as $c \rightarrow \infty$, which can be understood as resulting from $2M$ non-zero eigenvalues of $Z$ each of which is of magnitude $\simeq 1/M$. Similarly, as $c\to0$, the trace distance  goes to zero like $\simeq \frac{4 \sqrt{2} \sqrt{c}}{3 \pi }$. Comparing this leading behavior with  that  of  the distance to the maximally mixed state, $d_{\mathrm{tr}}(\rho_1,\mathbb{I}/N)$, which can be obtained from the first moment of the Mar\v{c}enko-Pastur law \eqref{marcenkopastur}, we find that $d_{\mathrm{tr}}(\rho_1,\mathbb{I}/N) \simeq \frac{4  \sqrt{c}}{3 \pi } \simeq 0.42 /\sqrt{NM}$, so that asymptotically $d_{\mathrm{tr}}(\rho_1,\rho_2) = \sqrt{2}\, d_{\mathrm{tr}}(\rho_1,\mathbb{I}/N)$ for $c \ll 1$, which is the same relation obeyed by the operator norm distance.

\section{The Exact Joint Eigenvalue Density}\label{section_exact}
We then proceed with the proofs of  Theorems \ref{derivativethm} and \ref{exact_theorem}. As mentioned earlier, despite the fact that Theorem \ref{derivativethm} is a known result, it is not widely known within the RMT community. We have therefore chosen to prove it here employing random matrix theory techniques.

\subsection{Proof of Theorem \ref{derivativethm} (Derivative Principle):}


 Consider an ensemble of random $N \times N$  Hermitian matrices $Z$ that is known to have a unitarily-invariant PDF  $P(Z)$; that is,
\begin{equation}
P(Z) dZ = P(Z')dZ', \qquad Z' = U Z U^\dagger,
\end{equation} for any unitary matrix $U$, where $dZ$ is the standard volume element for Hermitian matrices
\begin{equation}
dZ = \prod_{i \leq j} dZ^R_{ij}\prod_{i < j} dZ^I_{ij},
\label{hermvolel}
\end{equation}
and $Z^R_{ij}$ and $Z^I_{ij}$ are the real and imaginary parts of $Z_{ij}$ (with independent variables  $Z_{ij}^R$ for $i \leq j$ and $Z_{ij}^I$ for $i < j$).  Now let $\Phi_Z(K)$ be the matrix Fourier transform of $P(Z)$,
\begin{equation}
\Phi_Z(K)=\int  e^{-i\Tr\left[K Z \right] }\, P(Z) dZ
\end{equation}
where henceforth we shall assume that $K$ is also a Hermitian $N \times N$ matrix. From the unitary invariance of $P(Z)$, it follows that $\Phi_Z(K)$ is a symmetric function of the eigenvalues $\kappa_i$ of $K$; hence, we shall write the characteristic function as $\Phi_Z(\vec{\kappa})$, where  $\vec{\kappa}\equiv(\kappa_1,\kappa_2,\ldots,\kappa_N)$. We then define $\Psi_Z( \vec{z})$ as the ordinary Fourier transform of $\Phi_Z(\vec{\kappa})$:
\begin{equation}
\Psi_Z( \vec{z}) = \frac{1}{ (2 \pi)^N } \int  e^{i \vec{\kappa}\cdot \vec{z} }\, \Phi_Z(\vec{\kappa})\, d^{N}\!\kappa\ .
\end{equation}
Let us now show that  $\Psi_Z( \vec{z})$ is in fact the PDF  for the diagonal elements $z_i \equiv Z_{ii}$ of $Z$. For this, note the identity
\begin{equation}
\delta(Z) = \frac{1}{2^N \pi^{N^2}}\int e^{i \Tr( K Z)}\, dK
\end{equation}
where $\delta(Z) = \prod_{i \leq j} \delta(Z^R_{ij})\prod_{i < j} \delta(Z^I_{ij}) $ and the matrix integral  will  be understood to be over the subspace of Hermitian matrices with respect to the volume element \eqref{hermvolel}. The PDF $P(Z)$ can then be expressed as the integral
\begin{equation}
P(Z) =\frac{1}{2^N \pi^{N^2}} \int e^{i\Tr\left[ K Z\right] }\Phi_Z(K) dK.
\end{equation}
Now break up $Z$ as $Z_{\parallel} + Z_{\perp}$, where $Z_{\parallel} = \mathrm{diag}(z_1, z_2, \ldots z_N)$ and $Z_{\perp}$ the off-diagonal part of $Z$. The  marginal PDF for the diagonal elements is then given by
\begin{equation}
P(\vec{z}) = \int  dZ_\perp P(Z) =
 \frac{1}{2^N \pi^{N^2}} \int dZ_\perp \int e^{i \sum_{j=1}^{N} z_j K_{jj} +  i \Tr(K Z_\perp)} \Phi_Z(K) dK,
\end{equation}
where $dZ_{\perp} =  \prod_{i < j} dZ^R_{ij}\prod_{i < j} dZ^I_{ij}$. Exchanging the order of integration, we use the fact that
\begin{equation}
\int  e^{ i \Tr(K Z_\perp)} dZ_\perp = \pi^{ N(N-1) } \delta(K_\perp),
\end{equation}
where $K_\perp$ is the off-diagonal part of the matrix $K$. Hence,
\begin{equation}
\label{pdiag}
P(\vec{z}) =
 \frac{1}{(2 \pi)^N } \int e^{ i \sum_{i=1}^N z_i K_{ii}} \Phi_Z(\vec{\kappa}= \mathrm{spec}(K_\parallel))\prod_{i} dK_{ii}  ,
\end{equation}
where $K_{\parallel}$ is $\mathrm{diag}(K_{11},K_{22},\ldots,K_{NN})$. However, if $K$ is diagonal, the $K_{ii}$ are also its eigenvalues. Hence  $P(\vec{z})$ is the Fourier transform of $\Phi_Z(\vec{\kappa})$ and is therefore equal to  $\Psi_Z(\vec{z})$.

Now, since $Z$ is Hermitian, we can parametrize it as $Z= U \Lambda_Z U^\dagger$, where $U$ is unitary and $\Lambda_Z$ is the diagonal matrix of the eigenvalues $\vec{\lambda} = (\lambda_1,\lambda_2,\ldots,\lambda_N)$ of $Z$. A standard result from Random Matrix Theory  \cite{mehta_random_2004} states the volume element \eqref{hermvolel} can be  written as
\begin{equation}\label{measrel}
dZ = \frac{\pi^{ N(N-1)/2}}{\prod_{p=1}^{N} p!} \Delta^2(\vec{\lambda})\, \left(\prod_{i=1}^{N}d\lambda_i\right)\, D U,
\end{equation}
where $DU$ is shorthand for the normalized Haar measure on the group of $N \times N$ unitary matrices. From this it follows that if
$P(Z)$ is unitarily invariant,  the eigenvalue PDF $\density(\vec{\lambda})$ can be written as
\begin{equation}\label{jedpz}
\density( \vec{\lambda} ) = \frac{\pi^{ N(N-1)/2}}{\prod\limits_{p=1}^{N} p!} \Delta^2(\vec{\lambda}) P(\Lambda_Z) .
\end{equation}
Using Eqs. \eqref{pdiag} and \eqref{jedpz}, and similarly applying \eqref{measrel} to the measure $dK$, we can then express  $\density( \vec{\lambda} ) $ as
\begin{equation}\label{eqmu}
\density( \vec{\lambda}) =\frac{\Delta^2( \vec{\lambda})}{(2 \pi )^N \left( \prod\limits_{p=1}^{N} p! \right)^2 }  \int\Delta^2\left( \vec{\kappa}\right) \left\langle e^{i\Tr\left[ U \Lambda_K U^\dagger \Lambda_Z  \right] }\right\rangle_U  \Phi_Z(\vec{\kappa})\, d^N\!\kappa\, ,
\end{equation}
where $\Lambda_K$ is the diagonal matrix of the eigenvalues of $K$ and
\begin{equation}
\left\langle e^{i\Tr\left[ U \Lambda_K U^\dagger \Lambda_Z  \right] }\right\rangle_U \equiv \int  e^{i\Tr\left[U \Lambda_K U^\dagger   \Lambda_Z \right] }\, DU\, .
\end{equation}
This is the
well-known Harish-Chandra-Itzykson-Züber integral  \cite{itzykson_planar_1980}:
\begin{equation}\label{zuber1}
\left\langle e^{i\Tr\left[ U \Lambda_K U^\dagger \Lambda_Z \right] }\right\rangle_U=i^{-N(N-1)/2}\left(\prod\limits_{p=1}^{N-1}p!\right)\dfrac{\det(\exp[i \kappa_j\lambda_k]_{1\leq j,k\leq N})}{\Delta(\vec{\kappa})\Delta(\vec{\lambda})}.
\end{equation}
Inside the integral \eqref{eqmu}, we can use permutational symmetry of the integrand to replace   $\det(\exp[i \kappa_j\lambda_k]_{1\leq j,k\leq d})$ by $N! \exp[i\vec{\kappa}\cdot\vec{\lambda}]$, to obtain
\begin{equation}\label{nosecomo}
\density( \vec{\lambda}) = \frac{ i^{- N(N-1)/2} }{(2 \pi)^N  \left(\prod\limits_{p=1}^{N}p!\right) }\Delta( \vec{\lambda}) \int \Delta\left( \vec{\kappa}\right)e^{i\vec{\kappa}\cdot\vec{\lambda}} \,  \Phi_Z(\vec{\kappa})\,d^N\!\kappa\, .
\end{equation}
Finally, replacing the  arguments $\kappa_i$ in $\Delta(\vec{\kappa})$ by the partial derivative operators $-i\frac{\partial}{\partial \lambda_i}$ acting on $e^{i\vec{\kappa}\cdot\vec{\lambda}}$, we obtain relation \eqref{distribucion} between $\density(\vec{\lambda})$ and $\Psi_Z(\vec{\lambda})$, since $\Psi$ is the Fourier transform of $\Phi$.

\subsection{ Proof of Theorem \ref{exact_theorem}: Computation of $\Psi_Z(\vec{z})$ }\label{Psi_subs}

Next we turn to the derivation of \eqref{diagdist1} in Theorem \ref{exact_theorem}. First, suppose $Z = X + Y$, where   $X$ and $Y$ are independently drawn from unitarily invariant ensembles. Then, the PDF of the diagonal elements of $Z$ is simply the convolution of  $\Psi_X$ and $\Psi_Y$. Specializing this result to the problem of interest, we take $X=\rho_{1}$ and $Y= -\rho_{2}$, where $\rho_1$ and $\rho_2$ are sampled independently from the FTWL ensemble, with eigenvalue PDF  \cite{sommers_statistical_2004}
\begin{equation}\label{wishartdist}
\density(\vec{\lambda})=\dfrac{\Gamma(M N )}{\prod\limits_{j=0}^{N-1}\Gamma(M\!-\!j)\Gamma(N\!-\!j\!+\!1)} \delta_S\left(\vec{\lambda}\right)\prod_{i=1}^{N} \lambda_i^{M-N}\prod_{i<j}(\lambda_i-\lambda_j)^2,
\end{equation}
 where $\delta_S(\vec{\lambda})=\delta\left(\sum_{i=1}^{N}{\lambda_i}-1\right)I_S(\vec{\lambda})$ and $I_S(\vec{\lambda})$ is the indicator function on the probability simplex $S$,
\begin{equation}
S= \left\{ (\lambda_1, \lambda_2, \ldots \lambda_N) \left|{} \  \lambda_i \geq 0 , \sum_{i=1}^{N} \lambda_i =1\right. \right \}\, .
\end{equation}
In the FTWL ensemble, a random reduced density matrix $X = \rho_1$ can be expressed as
\begin{equation}
X = \frac{G G^\dagger}{\mathrm{Tr}(G G^\dagger)},
\end{equation}
where $G$ is an $N \times M$ matrix with independent Gaussian complex entries. It is then straightforward to show that the PDF of the diagonal elements is the symmetric  Dirichlet distribution
\begin{equation}\label{psiFTWL}
\Psi_X( \vec{x} ) = \frac{ \Gamma(MN) }{\Gamma(M)^N } \delta_S\left(\vec{x}\right)  \prod_{i=1}^{N} x_i^{M-1}  \, .
\end{equation}
Next, recalling that $Y=-\rho_{2}$, we have
$
\Psi_Y(\vec{x})=\Psi_X(-\vec{x});
$
hence,
\begin{equation}
\label{eqfinal}
\Psi_Z(\vec{z} ) = \int\limits_{\mathbb{R}^{N}}\!d^N\!x\, \Psi_X(\vec{x})\Psi_X(\vec{x}-\vec{z}).
\end{equation}
The region of integration, as well as the region of support of $\Psi_Z$, is set by the support conditions  of the $\Psi_X$ in the integrand.
The integration region for integral \eqref{eqfinal} is the one that lies in the intersection between the probability simplex $S$ and the shifted simplex
\begin{equation}
S' = \left \{ (x_1 + z_1,x_2 + z_2,\ldots, x_N + z_N)\left|{} x_i \geq 0 , \sum_{i=1}^{N}x_i =1\right. \right \}\, .
\end{equation}
We will call this region $T = S \cap S'$.
Since the diagonal elements of a density matrix lie in the  probability simplex, the diagonal elements of the difference between two random density matrices must lie in the  Minkowski difference set $R = S - S$, or explicitly,
\begin{equation}
R = \left\{ \vec{r} - \vec{r}\,'  |\vec{r},\vec{r}\,' \in S \right \} \, ,
\end{equation}
which is the $(N-1)$-dimensional convex polytope on the hyperplane $\{ (x_1 \ldots, x_N) \in \mathbb{R}^N| \sum_{i=0}^N x_i = 0\}$ with vertices at the points $\vec{e}_i - \vec{e}_j $ for $i\neq j$, where $\vec{e}_i$ are the standard unit vectors in $\mathbb{R}^N$. The condition $\vec{z} \in R$ is precisely the condition such that $ S \cap S'\neq \emptyset$ and hence the support condition for $\Psi_{Z}$.

To further characterize the integration region $T$,
let $P_k$ be the facet of $S$ with vertices $\left\lbrace \vec{e}_i | i \neq k\right\rbrace  $ and likewise $P'_k$ be the facet of $S'$ with vertices $\left\lbrace \vec{e}_i + \vec{z} | i \neq k\right\rbrace  $, with $\vec{z} \in R$. If $\vec{r}$ is a point in $T$ then the distances $d_k(\vec{r}\,)$ and $d'_k(\vec{r}\,)$ of $r$ to $P_k$ and $P_k'$ respectively, are given by
\begin{equation}
d_k(\vec{r}\,) = \sqrt{\frac{N}{N-1}}x_k \, \ \  , \ \   d'_k(\vec{r}\,) = \sqrt{\frac{N}{N-1}}(x_k - z_k)\, ,
\end{equation}
where $(x_k)_{k=1}^{N}$ are the coordinates of $\vec{r}$ (with $\sum_{i =1}^{N} x_i =1$). This implies that if $z_k > 0$, then the facet $P'_k$ is closer to $\vec{r}$ than the facet $P_k$, and conversely, if $z_k < 0$ then the facet $P_k$ is closer to $\vec{r}$ than the facet $P'_k$. Hence, $T$ is the convex polytope bounded by the $N$ facets $F_k$ where
\begin{equation}
F_k = \left \{ \begin{array}{cc} P_k\, , & z_k \leq 0 \\ P'_k & z_k >0 \end{array} \right. .
\end{equation}
Let us then parameterize points in a facet $F_k$ by
\begin{equation}
\vec{r} = \sum_{i \neq k}\theta^{k}_i \vec{e}_i + s_k \vec{z} \, \ \ \ \sum_{i \neq k} \theta^k_i = 1 \, , \ \  \theta^k_i \geq 0,
\end{equation}
where $s_k =1$ if $z_k >0$ and $s_k=0$ if $z_k \leq 0$.  When two facets $F_k$ and $F_{k'}$ meet, the intersection is  described by the condition
\begin{equation}
 \sum_{i \neq k}\theta^{k}_i \vec{e}_i + s_k \vec{z}  = \sum_{i \neq k'}\theta^{k'}_i \vec{e}_i + s_{k'} \vec{z} .
\end{equation}
Using these constraints and solving for the coefficients $\theta^k_i$, the vertices $\left\lbrace \vec{r}_j\right\rbrace $ of the integration region can be shown to be given by
\begin{equation}
\label{verts}
\vec{r}_j = \vec{\beta} + \gamma \vec{e}_j \, ,
\end{equation}
where
\begin{equation}
\vec{\beta} = \sum_{j: z_j >0} z_j \vec{e}_j \, , \ \ \ \  \gamma = 1  -\sum_{j: z_j >0} z_j.
\end{equation}
Thus, the integration region $T$ is also a regular simplex, obtained from the standard simplex $S$ by a shift $\vec{\beta}$ and uniform rescaling by the factor $\gamma$. Note that the support condition $\vec{z} \in R$ implies  that $\sum_j z_j = 0$, and hence that on $R$, $\gamma$ can be written as a symmetric function of the $z_i$; namely,
\begin{equation}
\gamma = 1  - \frac{1}{2}\sum_{j=1}^{N} |z_j | .
\end{equation}
Introducing  the change of variables
$
x'_k = \frac{1}{\gamma}(x_k - s_k z_k)
$
to undo these transformations, and combining Eqs. \eqref{psiFTWL} and \eqref{eqfinal}, we arrive at an expression for $\Psi$ in terms of an integral on the standard probability simplex $S$:
\begin{equation}
\label{diagdistintegral}
\begin{split}
\Psi_Z(\vec{z})= &\delta_R\left(\vec{z}\right)\frac{\Gamma(MN)^2}{\Gamma(M)^{2N}}  \, \gamma^{N(2M-1)-1}\,\\
&\times \int_{\mathbb{R}^N} \! \! d^N\vec{x}\ \ \delta_S\left(\vec{x}\right)\prod_i^{N} x_i^{M-1} \left(x_i + \frac{|z_i|}{\gamma}\right)^{M-1}.
\end{split}
\end{equation}
Where $\delta_R(\vec{z})$ is defined as in Equation \ref{surface_delta}.
%
Expanding the terms $\left(x_i + \frac{|z_i|}{\gamma}\right)^{M-1}$ in the integrand, we can use the multinomial Beta function
\begin{equation}
\frac{\prod_{i=1}^{N} \Gamma(\alpha_{i} )}{\Gamma\left(\sum_{i=1}^N \alpha_i \right)} = \int_{\mathbb{R}^N} \! \! d^N\vec{x}\ \ \delta_S\left(\vec{x}\right)\prod_{i=1}^{N} x_i^{\alpha_i-1},
\end{equation}
to arrive at the expression
\begin{equation}
\begin{split}
\Psi_Z(\vec{z}) = &\delta_R\left(\vec{z}\right)\frac{\Gamma(MN)^2}{\Gamma(M)^{N}}\!\!\\
&\times\sum_{k_1, ...,k_N}\frac{\gamma^{N(2M-1)-1- \sum_i k_i}}{(N(2M-1)-1-\sum_{i}k_i)!}\prod_{i=1}^{N}\frac{(2(M-1)-k_i)!}{k_i!(M-1-k_i)!}|z_i|^{k_i},
\end{split}
\end{equation}
where it is understood that the $k_i$ run over all values such that the arguments in the factorials are non-negative. We can now use the fact that
\begin{equation}
\frac{\gamma^{\ell}}{\ell!} = \frac{1}{2 \pi i } \oint\! ds\, e^{\gamma s}\,  s^{ - \ell -1 },
\end{equation}
and noting that the resulting sums under the integral are expressible in terms of the associated Laguerre polynomials as
\begin{equation}
\sum_{k_i=0}^{M-1}\frac{(2(M-1)-k_i)!}{k_i!(M-1-k_i)!}(s|z_i|)^{k_i} = (M-1)! (-1)^{M-1} L_{M-1}^{1-2 M}(s|z_i|),
\end{equation}
we finally obtain expression \eqref{diagdist1}.

It is also worth noting that a closed-form expression for $\Psi_Z(\vec{z}) $ is possible in terms of  the
so-called Lauricella  generalized hypergeometric function of type $A$  \cite{exton_multiple_1976}, defined as
\begin{equation}
F_A^{(n)}\left( a;\substack{ b_1, \ldots , b_n \\ c_1, \ldots ,c_n}|x_1, \ldots, x_n\right)=\sum\limits_{i_1,\ldots,i_n}\dfrac{(a)_{i_1+\cdots+i_n}(b_1)_{i_1}\cdots(b_n)_{i_n}}{(c_1)_{i_1}\cdots(c_n)_{i_n}i_1!\cdots i_n!}x_1^{i_1}\cdots x_n^{i_n},
\end{equation}
where
$
(a)_i=\dfrac{(a+i-1)!}{(a-1)!}
$ is the rising factorial. Therefore, we can alternatively express $\Psi_Z(\vec{z}) $ as
\begin{equation}
\label{diagdist2}
\begin{split}
\Psi_Z(\vec{z}) =   & \delta_R\left(\vec{z}\right) \frac{\Gamma(MN)^2\ \Gamma(2M-1)^N}{\Gamma(M)^{2N} \ \Gamma(N(2M-1)) } \gamma^{N(2M-1)-1} \\
 & \times F_A^{(N)}\left(\!\! \substack{N(1-2M)+1}{};\substack{  1-M\ , \ldots ,\ 1-M  \\ 2(1-M), \ldots , 2(1-M) }\left |-\frac{|z_1|}{\gamma}, \ldots, -\frac{|z_N|}{\gamma} \right.\right).
\end{split}
\end{equation}

 \section{The Asymptotic Eigenvalue Density} \label{section_Asymptotic}

We next proceed with the proof of Theorem \ref{disttheorem} for the AED. The approach we follow in this section again exploits the fact that the random matrix  $Z = \rho_1 - \rho_2$ is the difference of two independent random matrices drawn from unitarily invariant ensembles. In the asymptotic limit, it is well-known that any two such matrices satisfy the so-called \emph{freeness condition} underpinning Voiculescu's theory of \emph{free probability} for non-commuting variables  \cite{voiculescu_free_1992}.  In a manner analogous to the case of the sum (or difference) of two independent commuting random variables,  the AED of $Z$ can be shown to be given by the so-called \emph{free convolution} of the AEDs of $\rho_1$ and $-\rho_2$.

 \subsection{Free Convolution and the AED}

First, let us  briefly review the  aspects of Free Probability  that will be relevant to our analysis; the interested reader is referred to Nica and Speicher's book  \cite{nica_lectures_2006} on the topic for further details. As mentioned earlier, the ensemble of  $N \times N$ partial density matrices $\rho = \mathrm{Tr}_{\mathcal{H}_M}(|\psi\rangle\langle \psi|)$ where $|\psi  \rangle $ is uniformly sampled  from  $\mathcal{H}_M \otimes \mathcal{H}_{N}$, is the unitarily invariant FTWL random matrix ensemble. In the asymptotic limit $N \rightarrow \infty$, $M\to\infty$ with $c$ fixed, the normalized traces of integer powers of $\rho$ converge, almost surely, to the moments $m_n$ of a well-defined spectral measure, \emph{the asymptotic eigenvalue density} (AED) $\density_{\rho}(\lambda)$  \cite{alsmeyer_asymptotic_2013}:
 \begin{equation}
 m_n = \lim_{N\rightarrow \infty} \frac{1}{N} \mathrm{Tr}(\rho_N^{n}) = \int \density_\rho(\lambda) d\lambda \ \lambda^{n}.
 \end{equation}
  The AED for the FTWL ensemble was obtained in the context of random partial density matrices by Page  \cite{page_average_1993} following the Coulomb gas analogy. With the knowledge of $\density(\lambda)$, the theory of free probability allows us to calculate the AED for the sum  (or difference) of two random matrices $\rho_1$ and $\rho_2$ sampled from the FTWL ensemble.

 Free probability is a probability theory for non-commutative random variables satisfying a generalization of the ordinary notion of independence. Specializing to the case of random matrices, we say that two random matrices $A$ and  $B$ are said to be free if for all $k, m_i, n_i  \in \mathbb{N}_+$,
 \begin{equation}
 \left \langle \mathrm{Tr} \left( \Delta A^{m_1}\Delta B^{n_1}\Delta A^{m_2}\Delta B^{n_2} \ldots \Delta A^{m_k}\Delta B^{n_k}\right) \right \rangle = 0,
 \end{equation}
 where $\Delta X \equiv X - \langle \mathrm{Tr}( X)\rangle$. A well-known theorem of Voiculescu  \cite{voiculescu_limit_1991,nica_lectures_2006} states that if $(A_N)_{N\in\mathbb{N}}$ and $(B_N)_{N\in\mathbb{N}}$ are sequences of $N\times N$ matrices such that their asymptotic eigenvalue densities $\density_A(\lambda)$ and $\density_B(\lambda)$ exist, and if  $(U_N)_{N\in\mathbb{N}}$ is a sequence of Haar unitary $N\times N$ random matrices, then $U_NA_NU_N^\dagger$ and $B_N$ are asymptotically free as $N\rightarrow\infty$.
 From this theorem it follows that any two random matrices, each independently sampled from a unitarily-invariant ensemble, are asymptotically free.

 If $A$ and $B$ are free random variables with AEDs  $\density_A(\lambda)$ and $\density_B(\lambda)$,  the AED of the sum $C = A + B$ satisfies a generalized notion of convolution, known as \emph{free convolution}, and denoted by $\density_C(\lambda) = \density_A(\lambda) \boxplus \density_B(\lambda)$.
 For a given AED $\density(\lambda)$, it is convenient to define its Cauchy transform
 \begin{equation}
 G_{\density}(z)=\int_\mathbb{R}\dfrac{\density(\lambda)\, d\lambda\, }{z-\lambda},
 \end{equation}
  which is analytic on $\mathbb{C}\backslash \mathbb{R}$. In free probability theory, this function plays the role of a moment-generating (or characteristic) function. Indeed, the moments $m_n$ can be recovered from the Laurent expansion of $G_{\density}(z)$ for sufficiently large $|z|$:
 \begin{equation}
  G_{\density}(z)= \sum_{n=0}^{\infty}\dfrac{m_n}{z^{n+1}}\,  ,
  \label{Cauchtransf}
  \end{equation}
 and using the Stiltjes inversion formula  \cite{nica_lectures_2006}, the AED $\density(\lambda)$ can be recovered from $G_{\density}(z)$ according to:
  \begin{equation}\label{inversion}
 \density(\lambda)=-\dfrac{1}{\pi}\lim\limits_{\epsilon\rightarrow 0}\mathrm{Im} G_{\density}(\lambda+ i \epsilon).
 \end{equation}
 Closely related to the moments $m_n$ are the so-called \emph{free cumulants} $k_{n}$, which in analogy with ordinary cumulants are those combinations of the moments $m_n$ that are additive under the sum of two free random variables. The free cumulant generating function is the so-called $\mathcal{R}$ transform, which is connected to the Cauchy transform through the functional equations
 \begin{equation}
 G_{\density}\left(\mathcal{R}_{\density}+\dfrac{1}{z}\right)=z,
 \end{equation}
 and
 \begin{equation}\label{relationrg}
 \mathcal{R}_{\density}(G_{\density}(z))+\frac{1}{G_{\density}(z)}=z.
 \end{equation}
 The free cumulants are obtained from  the power series of $ \mathcal{R}_{\density}(z)$ about the origin:
 \begin{equation}\label{rtransform}
 \mathcal{R}_{\density}(z):=\sum\limits_{n=0}^{\infty}k_{n+1}z^n.
 \end{equation}
 For the sum of two free random variables, additivity of the free cumulants implies additivity of the respective $\mathcal{R}$ transforms:
  \begin{equation}\label{rtransformsum}
 \mathcal{R}_{\density_A\boxplus\density_B}(z)=\mathcal{R}_{\density_A}(z)+\mathcal{R}_{\density_B} (z).
 \end{equation}
 Thus, the AED of the sum of two free random variables can be obtained using \eqref{inversion} once the Cauchy transform is retrieved from $\mathcal{R}_{\density_A\boxplus\density_B}(z)$ by means of \eqref{relationrg}.

 \subsection{Proof of Theorem \ref{disttheorem} (AED for $\rho_1-\rho_2$): }
  \label{sectiondiff}
 We now specialize the previous discussion to the case of interest. For a random $N\times N$ partial density matrix $\rho$ playing the role of either $\rho_1$ or $\rho_2$, the AED was first computed by Page  \cite{page_average_1993} (see also the result from C. Nadal et al.  \cite{nadal_statistical_2011}). Using the well-known 2D Coulomb gas analogy, he showed that for $c=N/M<1$ the rescaled AED is
 \begin{equation}\label{rotilde}
 \tilde{\density}(x)=\dfrac{1}{2\pi cx}\sqrt{x-x_-}\sqrt{x_+ -x},
 \end{equation}
 with $x_\pm=c(1\pm c^{-1/2})^2$.
 Using the fact that the non-zero eigenvalues of the partial density matrices of a bipartite pure state are equal, it is not difficult to show that this result extends to the result of equation \eqref{marcenkopastur} for all values of $c$. As pointed out by Nechita  \cite{nechita_asymptotics_2007}, this AED corresponds to the well-known Mar\v{c}enko-Pastur law that arise from the infinite free Poisson process  \cite{nica_lectures_2006}. In terms of the variable $c$, the free cumulants for the Mar\v{c}enko-Pastur  distribution \eqref{marcenkopastur} are well-known and given by $k_n=c^{n-1}$ (see e.g., \cite{nica_lectures_2006}) . Using \eqref{rtransform}, the expression for the $\mathcal{R}$ transform is
 \begin{equation}\label{r1}
 	 \mathcal{R}_{\tilde{\density}}(z)=\sum\limits_{n=0}^{\infty}k_{n+1}z^n=\sum\limits_{n=0}^{\infty}c^{n}z^n=\dfrac{1}{1-cz}.
 \end{equation}
 The rescaled AED  of $-\rho$ is given by the reflected density $\tilde{\density}(-x)$ with free cumulants $k_n=(-1)^nc^{n-1}$, and  hence $\mathcal{R}$ transform $\sum\limits_{n=0}^{\infty}(-1)^{n+1}c^{n}z^n= -\dfrac{1}{1+cz}$.
 We therefore have, from \eqref{rtransformsum}, that the $\mathcal{R}$ transform of the rescaled AED of $Z=\rho_1-\rho_2$ is
 \begin{equation}
 	\mathcal{R}(z)=\dfrac{1}{1-cz}+\left(-\dfrac{1}{1+cz}\right)=\dfrac{2cz}{1-c^2z^2}.
 \end{equation}
 Making use of relation \eqref{relationrg}, a cubic equation for $G(z)$ can be obtained:
 \begin{equation}\label{cubicg}
 	\dfrac{2cG(z)}{1-c^2G(z)^2}+\dfrac{1}{G(z)}=z.
 \end{equation}
 The three roots for $G(z)$, indexed by $k=0,1,2$, can be given in terms of trigonometric functions as
 	\begin{equation}\label{gk}
 		 G^{(k)}=\dfrac{2\sqrt{(2-c)^2+3z^2}}{3cz}\sin\left[\frac{1}{3}\text{Arcsin}\left(\eta(z,c)\right)+k\frac{2\pi}{3}\right]+\dfrac{c-2}{3cz},
 	\end{equation}
 	where
 		\begin{equation}
 		\eta(z)\equiv \frac{9(c+1)z^2+(2-c)^3}{((2-c)^2+3z^2)^{3/2}},
 		\end{equation}
 	and $\text{Arcsin}(z)=-i\text{Log}(iz+(1-z^2)^1/2)$ and $\text{Log}(z)$ is the principal branch of the $\log(z)$ function with branch cut along the negative real axis. The actual function $G(z)$ is built piecewise from the roots $G^{(k)}$ in such a way that $G(z)$ is analytic in the region $\mathbb{C} \setminus \mathbb{R}$, and  decays as $1/z$ as $|z|\to\infty$. As mentioned previously, the solution depends on whether $c<2$ or $c>2$, so each case must be discussed separately.

\begin{figure}[H]
 \centering
         \begin{subfigure}{0.43 \textwidth}
         		\centering
         		\includegraphics[scale=0.25]{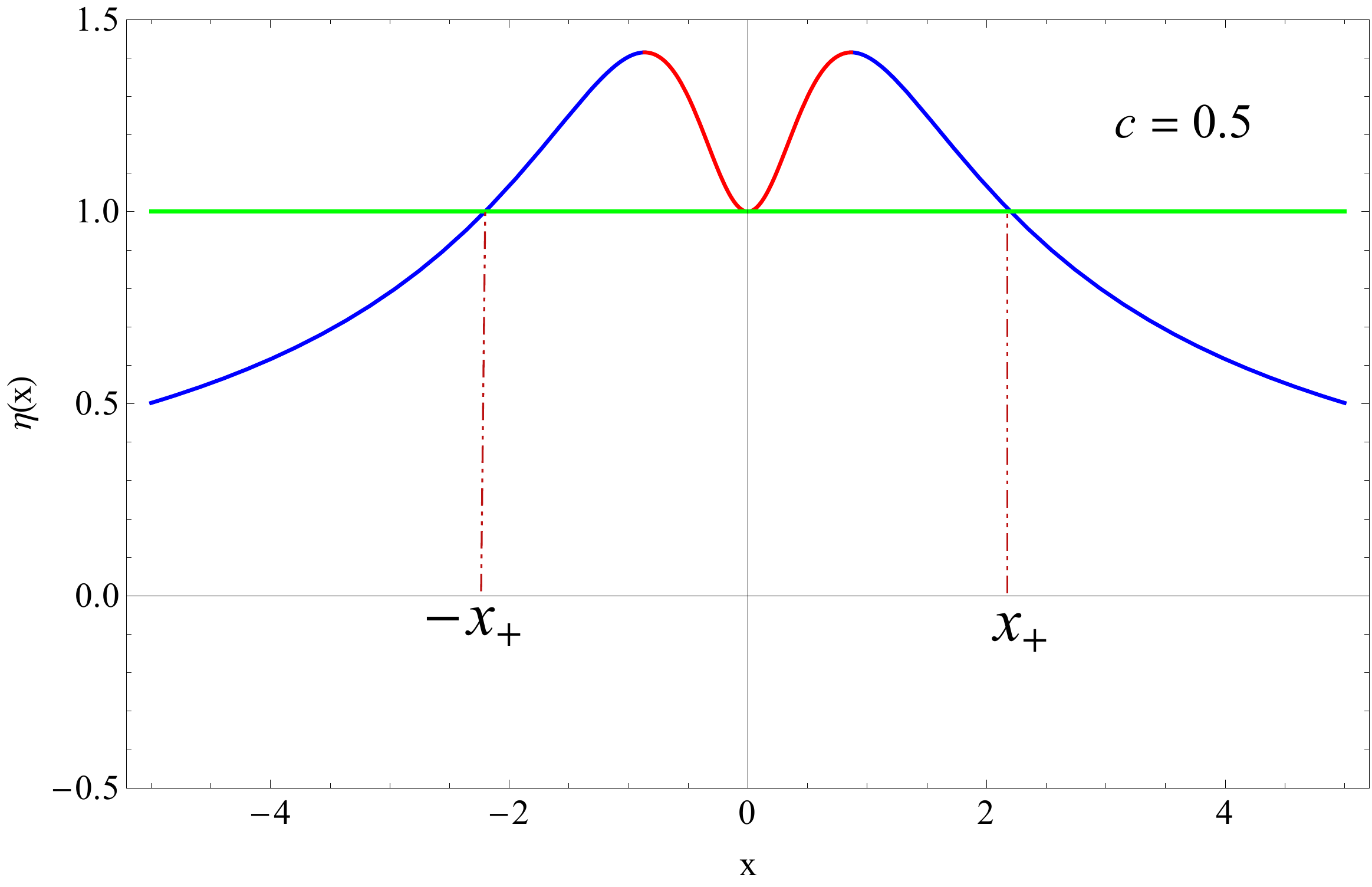}
         		\caption{$c<2$}
         		\label{cmenos2}
         \end{subfigure}
         \quad
          \begin{subfigure}{0.43\textwidth}
                         \centering
                         \includegraphics[scale=0.25]{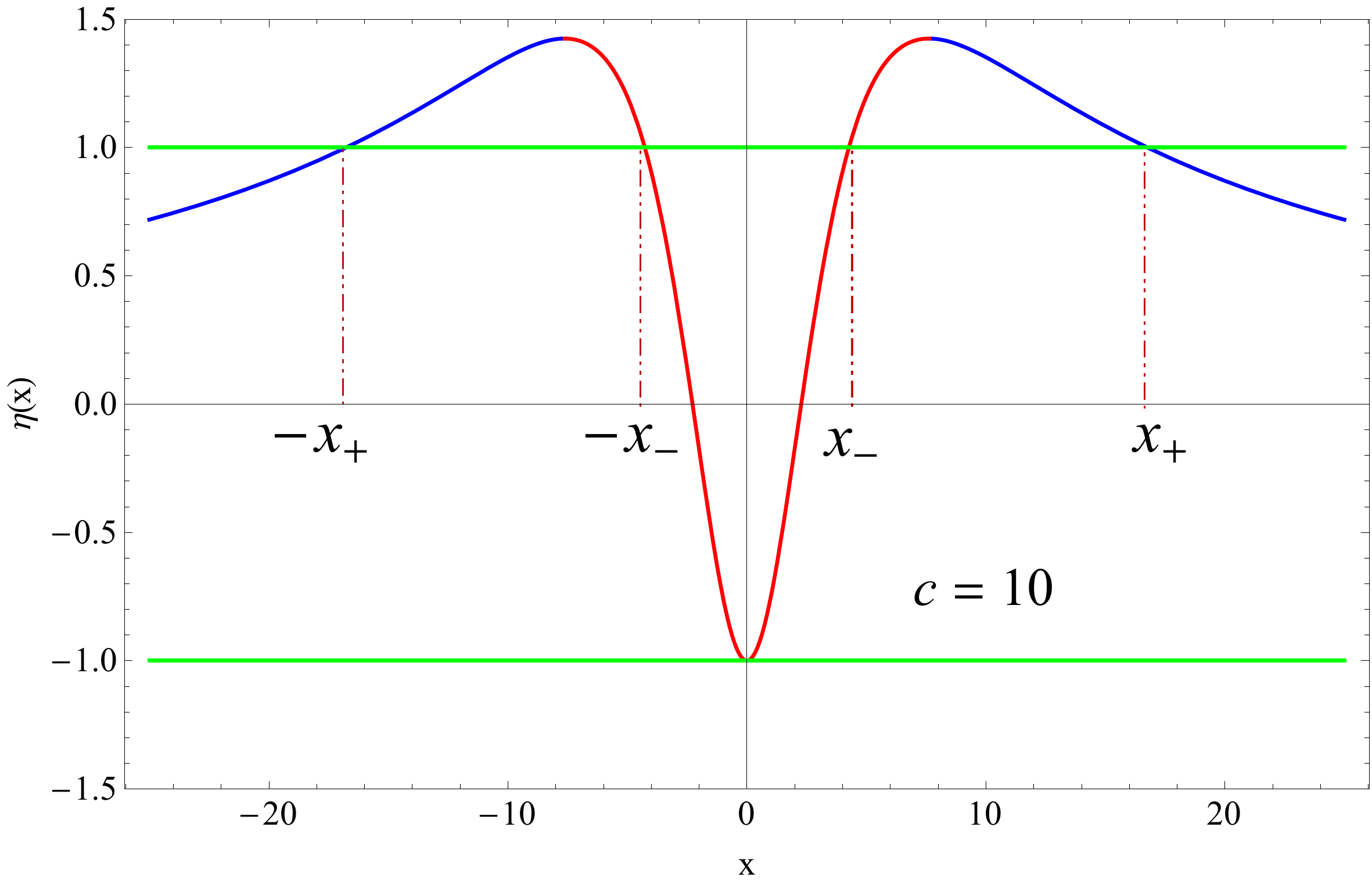}
                         \caption{$c>2$}
                         \label{cmas2}
           \end{subfigure}
           \caption{Plots of $\eta(x)$ for $c<2$ and $c>2$. Color of the curves indicates which root $G^{(k)}$ is used for $G(x+i \epsilon)$: blue for $G^{(0)}$ and red for $G^{(1)}$.}
 \label{etax}
 \end{figure}

As our interest is in the AED obtained via the Stieltjes inversion formula \eqref{inversion}, we only need to concentrate in obtaining the appropriate expression for $G(x+i\epsilon)$. In the limit $|x|\to\infty$, the root decaying as $1/x$ is $G^{(0)}$. To examine the behavior of this function in the real axis we can express the  $\text{Arcsin}$ in this root as
 \begin{equation}\label{eqArcSin}
   \text{Arcsin}(\eta(x+i\epsilon))=\left \{ \begin{array}{lcl}
  \frac{\pi}{2}+i\text{sgn}(\eta'(x))\log(\eta(x)+\sqrt{\eta^2(x)-1}) & {} & |\eta(x)| > 1 \\
  \arcsin(\eta(x)) & { } & |\eta(x)| \leq 1
  \end{array}
  \right. ,
   \end{equation}
 where $\text{sgn}(\cdot)$ is the sign function and $\arcsin(x)$ is the ordinary inverse sine function with domain $[-\pi/2,\pi/2]$. From \eqref{eqArcSin} we can see that a discontinuity arises when $\eta'(x)$ changes sign, which is indeed the case in both $c\leq 2$ and $c>2$ as  shown in Figs. \ref{cmenos2} and \ref{cmas2}, respectively. To analytically continue $G$, we must then switch to the root $G^{(1)}$ in the region $|x|<\frac{|2-c|\sqrt{c}}{\sqrt{c+1}}$ (see Fig. \ref{etax}).
 \noindent As a result we can express $G(x+i\epsilon,c)$ for all values of $x$ as
 \begin{equation}\label{gfinal}
  G(x+i\epsilon) = \frac{1}{(x+i\epsilon)}\left[f_1(x)-if_2(x)\right],
  \end{equation}
 where $f_1(x)$ and $f_2(x)$ are given by
\begin{equation}
 f_1(x)=\left \{ \begin{array}{lcl}
 \frac{\sqrt{(2-c)^2+3x^2}}{3c}\cosh\left(\frac{l(x)}{3}\right)+\frac{c-2}{3c} & {} & |\eta(x)| > 1 \\
 \frac{2\sqrt{(2-c)^2+3x^2}}{3c}\sin\left[\frac{1}{3}\arcsin\left(\eta(x,c)\right)+\frac{2\pi}{3}\right]+\frac{c-2}{3c} & { } & |\eta(x)| \leq 1
 \end{array}
 \right. ,
\end{equation}
 and
 \begin{equation}
  f_2(x)=\left \{ \begin{array}{lcl}
  \frac{\mathit{sgn}(x)\sqrt{(2-c)^2+3x^2}}{\sqrt{3}c}\sinh\left(\frac{l(x)}{3}\right) & {} & |\eta(x)| > 1 \\
  0 & { } & |\eta(x)| \leq 1
  \end{array}
  \right. ,
  \end{equation}
  and where $l(x)=\log\left(\eta(x)+\sqrt{\eta^2(x)-1}\right)$.
  Finally,  to obtain the PDF $\tilde{\density}(x)$ via the Stieltjes inversion formula \eqref{inversion}, we use the identity
  \begin{equation}\label{id_principal_v}
  \lim_{\epsilon\to 0^+}\frac{1}{x+i\epsilon}=\mathcal{P}\left( \frac{1}{x}\right) - i\pi \delta(x),
  \end{equation}
  where $\mathcal{P}\left(\cdot\right)$ is the principal value of a function. Explicitly,
  \begin{equation}\label{eqrhof}
  \tilde{\density}(x) = -\frac{1}{\pi}\lim_{\epsilon\to 0^+ }\mathrm{Im}G(x+i\epsilon) = \mathcal{P}\left(\frac{1}{x}\right)\frac{f_2(x)}{\pi}+\delta(x)f_1(0).
  \end{equation}
  From equation \eqref{eqrhof} we can obtain the results from Theorem \ref{disttheorem}.  For $c<2$ we have that $\eta(0) = 1$, therefore $f_1(0) =0 $, thus there is no point distribution for $c<2$. The non-zero region in \eqref{roasymptotic1} arises where $f_2(x)\not= 0$ $(\eta(x)>1)$, which occurs when $|x|\leq x_+$. For $c \leq 2$ we have that  $\eta(0)=-1$, so $f_1(0)=\frac{c-2}{c}$ thus obtaining the point distribution at $x=0$. The non-zero region in \eqref{roasymptotic2} arises where $f_2(x)\not= 0$ $(\eta(x)>1)$, that is for $x_-\leq |x|\leq x_+$.
 \paragraph{}

To conclude this section, it is worth mentioning that our calculation in the asymptotic regime is also related to the works of Auburn, Banica and Nechita  \cite{aubrun_partial_2010,banica_asymptotic_2013} where they calculate the eigenvalue PDF for the partial transpose $W^{PT}$ of a random Wishart matrix $W\in M_{dn}(\mathbb{C})\simeq M_{d}(\mathbb{C})\otimes M_{n}(\mathbb{C})$ of parameters $(dn,dm)$.
As shown in  \cite{banica_asymptotic_2013}, in the limit  $d\to\infty$, the spectral distribution of $mW^{PT}$ converges in moments to a free difference of {free Poisson} distributions with parameters $m(n\pm1)/2$. In the light of this approach, we can see our PDF as the formal limit $m=2c/n\to 0$, where $c$ is the fixed ratio of the Hilbert space dimensions. In that case, the spectral distribution of $mW^{PT}$ converges to the free difference of \emph{free Poisson} distributions with the same parameter $c$ which corresponds to the main result of this section.
%

\section{Moments and the distance between two random states}\label{section_Moments}

Finally, we turn to the proof of Theorem \ref{momentsth}. To obtain the general expression for the function $m_z$ in the theorem, we first compute the ordinary moments of the AED, which can be obtained from the Laurent expansion of the Cauchy transform $G_\varrho(z)$ \eqref{Cauchtransf};  from the symmetry of the AED it is clear that only even moments exist. The absolute complex moments can the be obtained by means of Carlson's theorem, which allows us to analytically continue the function for the discrete moments.

\subsection{Calculating the moments $m_n$ for even $n$}	

We will begin by calculating the moments $m_n = \left<x^n\right>$, for integer $n$. The moments can be read off from the Laurent expansion $G_\varrho(z)$ as $z \rightarrow \infty:$ \begin{equation}\label{momentsG}
G_{\density}(z) = \sum_{n=0}^{\infty} \frac{m_n}{z^{n+1}}.
\end{equation}
Now, from equation \eqref{cubicg} and with the change of variable $z = \frac{1}{\epsilon}$ we obtain:
\begin{equation}
\frac{G}{\Theta(G)}=\epsilon,
\end{equation}
where
\begin{equation}
\Theta(G) = \frac{c^2G^2-2cG^2-1}{c^2G^2-1}.
\end{equation}
This particular form is useful as it allows us to obtain $G$ as a power series of $\epsilon$ by applying the
Lagrange-B\"{u}rmann's inversion formula  \cite{lagrange_nouvelle_1770, burmann_essai_1799}; namely,
\begin{equation}\label{lagrangeburmann}
G(z) = \sum_{n=1}^{\infty} \frac{1}{n}\left( \frac{1}{(n-1)!}\lim_{G\rightarrow 0}\left( \frac{d^{n-1}}{dG^{n-1}}\Theta(G)^n\right)\right) \epsilon^n.
\end{equation}
The coefficients of the series in \eqref{lagrangeburmann} will be the moments $m_{n-1}$ according to \eqref{momentsG}. Thus, the moments are
\begin{equation}
m_{n-1} = \frac{1}{n}\left[G^{n-1}\right]\Theta(G)^n,
\end{equation}
where $[G^n]$ is an operator that extracts the coefficient of $n$-th power of $G$ on the Taylor series expansion of a function. Expanding $\Theta(G)^n$ as a double sum we get
\begin{equation}
\Theta(G)^n = \sum_{j,k=0}^{\infty} \frac{n(n+j-1)!}{(n-k)!k!}c^{2j+k}(2-c)^k G^{2(k+j)},
\end{equation}
and applying the $[G^{n-1}]$ operator we obtain:
\begin{equation}
m_{n-1} = \sum_{j=0}^{\infty} \frac{(n+j-1)!c^{n-k-1}(2-c)^{\frac{n-1}{2}-j}}{j!\left(\frac{n+1}{2}+j\right)!\left(\frac{n-1}{2}+j\right)!(n-1)!}.
\end{equation}
From the extraction of the $(n-1)$th power of $G$ we have that $n-1=2(k+j)$, and so $n-1$ has to be even. Now, letting $l=\frac{n-1}{2}$, we obtain
\begin{equation}
m_{2l} = c^l(2-c)^l\sum_{j=0}^{\infty}\frac{(2m+j)!}{j!(m+1+j)!(m-j)!}\left(\frac{c}{2-c}\right)^j,
\end{equation}
which can be written as a Gauss hypergeometric function ${}_2F_1(a,b;c;z)$ as follows:
\begin{equation}\label{momentos}
m_{2l} = \frac{(2l)!}{l!(l+1)!}c^l(2-c)^l {}_2F_1\left(2l+1, -l; l+2; \frac{c}{c-2}\right).
\end{equation}
To analytically extend to complex values of $l$, we will need expressions for ${}_2F_1(a,b;c;x)$ in which $|x| < 1$. For any integer $l >0$, we can use  standard transformations of the hypergeometric function on \eqref{momentos} to show that $m_{2l}(c) = \mu(l)$, where (for positive integer $l$ so far),
\begin{equation}\label{even2}
\mu(l) = \left\{ \begin{array}{ll} \dfrac{\Gamma(2 l+1) (2 c)^l }{\Gamma(l+1)\Gamma (l+2)}\, _2F_1\left(1-l,-l;l+2;\frac{c}{2}\right)& \text{for } c<2, \\
{} & {} \\ 2 c^{2 l-1} \, _2F_1\left(1-l,-2 l;2;\frac{2}{c}\right) & \text{for } c>2 .
\end{array}
\right.
\end{equation}
 Notice that if $\mu(l)$ is evaluated at $l=0$, it yields the result $2/c$ for $c>2$, which corresponds to the integral of $\tilde{\density}(x)$ excluding the Dirac atom at the origin.   In the next subsection we will show that the equality  $m_{2l} = \mu(l)$ extends to  complex values of  $l$ with $\text{Re}(l) > 0$.

\subsection{Proof of Theorem \ref{momentsth} (the  absolute moments $m_z$): }
To show that $m_{2l} = \mu(l)$  is indeed the correct expression for all  absolute moments with complex $l$ with positive real part, let us define $f(z)$ as
\begin{equation}\label{eqfz}
f(z)=m_{2z} -\mu(z), \qquad Re(z) \geq 0,
\end{equation}
where $m_{2z}$ is the moment function
\begin{equation}\label{mom_fun}
m_{2z} = 2\int\limits_{0^+}^{\infty}|x|^{2z}\tilde{\density}(x)dx ,
\end{equation}
and $\mu(z)$ is as defined in \eqref{even2}. The  lower limit $0^+$ of the integral \eqref{mom_fun} signifies that the integral is performed excluding the Dirac atom in $\tilde{\density}(x)$ at the origin for $c>2$. This choice guarantees that   $m_{2z}$ is analytic for all $\text{Re}(z) \geq 0$ and that $f(0) =0$ for all $c$. Moreover, from the analyticity  of the hypergeometric function with respect to its parameters, it is also ensured that $f(z)$ is analytic (and hence continuous) for all $\text{Re}(z) \geq 0$ (for $\text{Re}(z)<0$, we understand $f(z)$ to be the corresponding analytic extension). From the results of the previous section, we know that $f(z)=0$ for all  $z \in \mathbb{N}$. Using Carlson's theorem we can then show that in fact, $f(z)=0$ for all $z\in\mathbb{C}$. Since the integral in  \eqref{mom_fun}  is only valid in the region $\text{Re}(z)\geq 0$, the vanishing of $f(z)$ in that region suffices for Theorem \ref{momentsth}.  Outside this region, the vanishing of $f(z)$ tells us that the analytic extension of the integral in \eqref{mom_fun} is given by $m_{2z}$.

Carlson's theorem states that if $f(z)$ is analytic in $\text{Re}(z)>0$, continuous in $\text{Re}(z)\geq 0$, and satisfies the conditions:
\begin{subequations}
\begin{eqnarray}
(1)& \qquad |f(z)|\leq Ce^{\tau |z|}, & Re(z) \geq 0,\ \tau,C<\infty, \label{carlson1}\\
(2)& \qquad |f(iy)|\leq Ce^{c|y|}, & c < \pi, \label{carlson2} \\
(3)&\qquad f(n)=0, & n\in \mathbb{N}_{>0}, \label{carlson3}
\end{eqnarray}
\end{subequations}
 then $f$ is identically zero  \cite{bailey_generalized_1935,rubel_necessary_1955}. Analyticity and continuity of $f(z)$ in the required domains was previously ascertained, as was condition (3).  Conditions (1) and (2) are satisfied if they can be verified separately for $m_{2z}$ and $\mu(z)$. Indeed, for $m_{2z}$,  we have $|m_{2 z}| \leq x_+^{ 2 Re(z)}$ and $|m_{2 (i y)}| \leq 1$.  To ascertain conditions (1) and (2) for $\mu(z)$,    the asymptotic behavior of the ${}_2F_1$ function for large parameters must be examined. These asymptotics are  covered in two papers by Paris \cite{paris_asymptotics_2013,paris_asymptotics_2013-1}. Using the expansions of that reference, as well as the asymptotics of the Gamma function, it is not difficult to show that for $|z|\rightarrow\infty$, expression  \eqref{even2} behaves asymptotically as
\begin{equation}
\mu(x) \approx \left\{ \begin{array}{ll}
\frac{(2 c)^z \left(1-\frac{c}{2}\right)^{3 z+1} f_1(t_1(c),c) e^{z \psi_1 (t_1(c),c)}}{(z+1) \sqrt{z} \sqrt{2 \pi  \psi_1''(t_1(c),c)}}& \text{for } c<2,\\
{} & {} \\
\frac{2 c^{2 z-1} \left(1-\frac{2}{c}\right)^{z-1}f_2(t_2(c),c) e^{z \psi_2 (t_2(c),c)}}{(2z+1)\sqrt{z} \sqrt{2 \pi  \psi_2 ''(t_2(c),c)} }& \text{for } c>2,
\end{array}
\right.
\end{equation}
where
\begin{subequations}
\begin{eqnarray}
t_1(c)&=&\frac{4}{1 +\sqrt{1+4 c}},\\
\psi_1(t,c)& =& 2 \log[t] - \log[t - 1] - 2 \log[1 - c/2 t],\\
f_1(t,c) &= & \frac{1}{\left(1-\frac{c}{2}t\right)^2},\\
t_2(c)&=&\frac{4 \left(1-\frac{c}{2}\right)}{3 \left(1-\sqrt{1-\frac{8}{9} \left(1-\frac{c}{2}\right)}\right)},\\
\psi_2(t,c)&=&2 \log\left[\frac{t}{t-1}\right]+\log\left[1-\frac{t}{1-\frac{c}{2}}\right],\\
f_2(t,c)&=&\frac{t}{(t-1) \left(1-\frac{t}{1-\frac{c}{2}}\right)}.
\end{eqnarray}
\end{subequations}
The asymptotic expansion acquires a particularly  simple form when expressed in terms of $x_+$, namely,
\begin{equation}
\mu(x) \approx \left\{ \begin{array}{ll}
\dfrac{x_+^{2z}}{(z+1)\sqrt{z}}\left(\left(1-\frac{c}{2}\right)\dfrac{f_1(t_1(c),c)}{\sqrt{2\pi\psi_1''(t_1(c),c)}}\right)& \text{for } c<2,\\
{} & {} \\
\dfrac{x_+^{2z}}{(2z+1)\sqrt{z}}\left(\left(1-\frac{2}{c}\right)\dfrac{f_2(t_2(c),c)}{\sqrt{2\pi\psi_2''(t_2(c),c)}}\right)& \text{for } c>2,
\end{array}
\right.
\end{equation}
which clearly are bounded by exponential functions of the form of \eqref{carlson1} and \eqref{carlson2}. Thus, by Carlson's theorem it follows that $f(z)=0$, and thus  for $Re(z) \geq 0$, $m_{2z} = \mu(z)$. Apart from a trivial change of variable ($2 z \rightarrow z$), Theorem \ref{momentsth} follows by noting that when $z$ is restricted to $Re(z) >0$, the lower limit of integral \eqref{mom_fun} can be replaced by $0$, so as to include the Dirac atom, with no effect whatsoever.

\subsection{The case $z=1$}
\label{absmom1}

For the asymptotic trace distance in  corollary \ref{corollary} we need to specialize to the case $z =1$ in Theorem \ref{momentsth}, in which case, 

\begin{equation}
m_1=\left\{
\begin{array}{ll}
\frac{8 \sqrt{2} \sqrt{c} \, _2F_1\left(-\frac{1}{2},\frac{1}{2};\frac{5}{2};\frac{c}{2}\right)}{3 \pi },&  c\leq 2,\\
{}& {}\\
\dfrac{2c-1}{c}, &c >2.
\end{array}
\right.
\label{hypergeo}
\end{equation}
To obtain the simplified expression \eqref{distance1} from \eqref{hypergeo}, we can make successive use of the recurrence relations
\begin{multline}
\, _2F_1(a,b;c;z)=\frac{((c-1) (c-2) (1-z)) \, _2F_1(a,b;c-2;z)}{z (a-c+1) (b-c+1)}\\
+\frac{((c-1) (-z (a+b-2 c+3)-c+2)) \, _2F_1(a,b;c-1;z)}{z (a-c+1) (b-c+1)},
\end{multline}
and
\begin{multline}
\, _2F_1(a,b;c;z)=\frac{(z (a-b-1)+2 b-c+2) \, _2F_1(a,b+1;c;z)}{b-c+1}\\
+\frac{((b+1) (z-1)) \, _2F_1(a,b+2;c;z)}{b-c+1},
\end{multline}
together with the identity
\begin{equation}\label{hypergeometric3}
\arcsin(z)={}_2F_1\left(\dfrac{1}{2},\dfrac{1}{2},\dfrac{3}{2},z^2\right)z.
\end{equation}
With these identities, it follows that
\begin{equation}
\, _2F_1\left(\frac{1}{2},-\frac{1}{2};\frac{5}{2};\frac{c}{2}\right)=\dfrac{3(2c-1)\arcsin\left(\sqrt{\frac{c}{2}}\right)}{4\sqrt{2}c^{3/2}}+\frac{3 \sqrt{1- \frac{c}{2}} (c+1)}{8 c},
\end{equation}
which yields the result of Corollary \ref{corollary} for $c<2$ when multiplied by  $\frac{8 \sqrt{2} \sqrt{c}}{6\pi}$.

%
%
%

\section{Conclusions}
The aim of this paper has been to quantify the distance between two random density matrices, each independently sampled from the so-called Fixed Trace Wishart-Laguerre (FTWL) Ensemble. To this end, we have  successfully obtained expressions for the joint PDF of the eigenvalues of the difference matrix given finite Hilbert space sizes,  the limiting eigenvalue density (AED) in the asymptotic limit of infinite Hilbert space dimensions, and closed-form expressions for the absolute moments of the asymptotic density. The asymptotic expressions were used to quantify the almost sure asymptotic behavior of two distance measures for the two random states; namely, the operator norm distance and the trace distance, both of which are of the same order as the corresponding distances to the maximally mixed state, as expected from the concentration of measure phenomenon of partial density matrices around the maximally mixed state  \cite{page_average_1993,popescu_entanglement_2006,hayden_aspects_2006}.

\section{Acknowledgements}
 The authors are grateful to Matthias Christandl, and Satya Mujumdar for helpful discussions. AB gratefully acknowledges funding by Uniandes, proyecto Ciencias B\'{a}sicas No. 114-2013.

\section{Appendix}
\subsection{Appendix A}
\begin{figure}[H]
	\centering

	\begin{tabular}{cc}
		\includegraphics[scale=0.28]{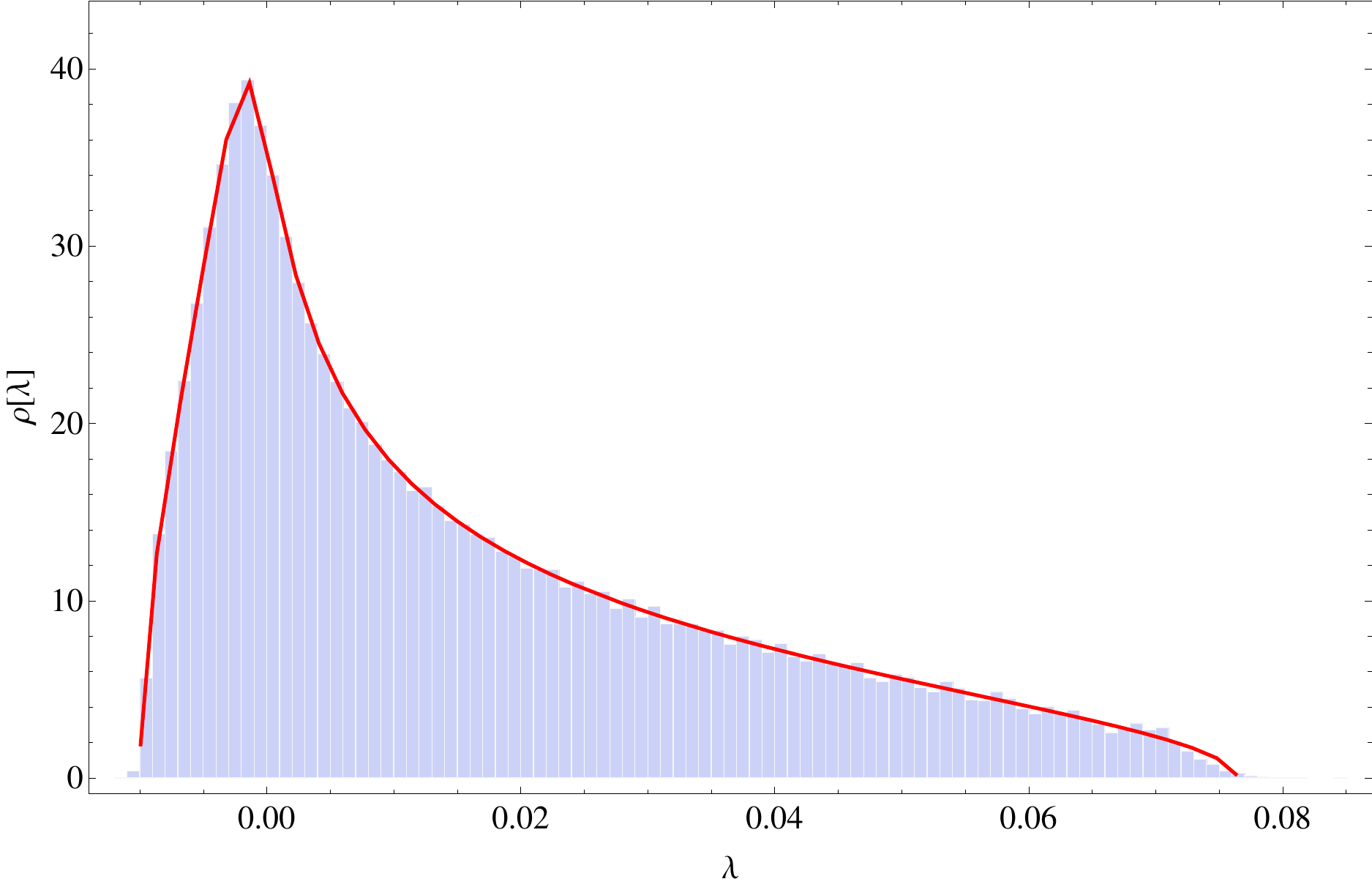}&\includegraphics[scale=0.28]{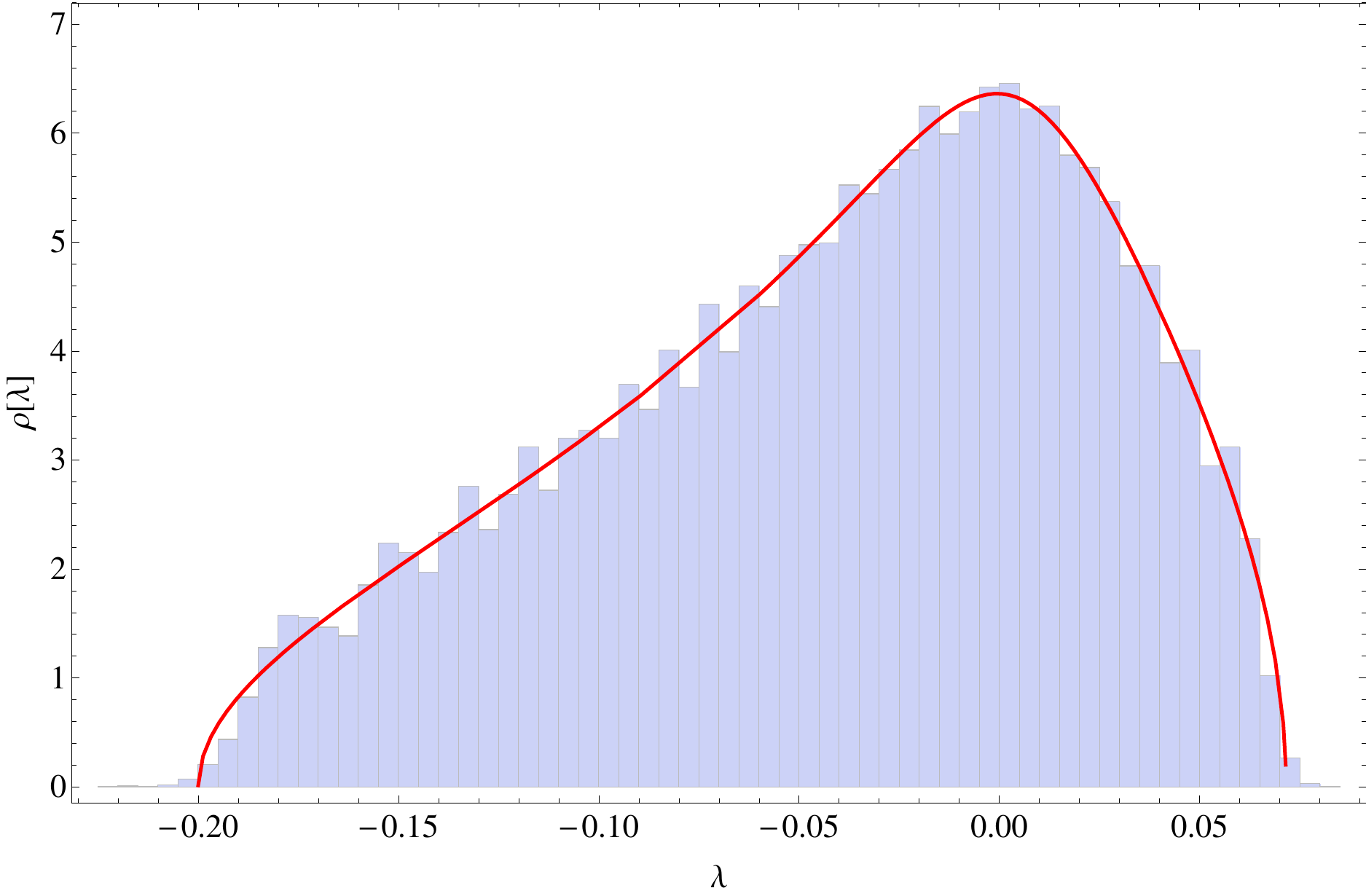}
	\end{tabular}

	\caption{Eigenvalue density: Theoretical results (red line) vs numerical results(blue bars) for the aymptotic case. (Left: $N=M=50$, $c=1$, $\eta=0.2$)(Right: $c=0.5$, $\eta=2$) }
	\label{compararion}
\end{figure}

\begin{figure}[H]
	\centering
	\begin{tabular}{cc}
		\includegraphics[scale=0.28]{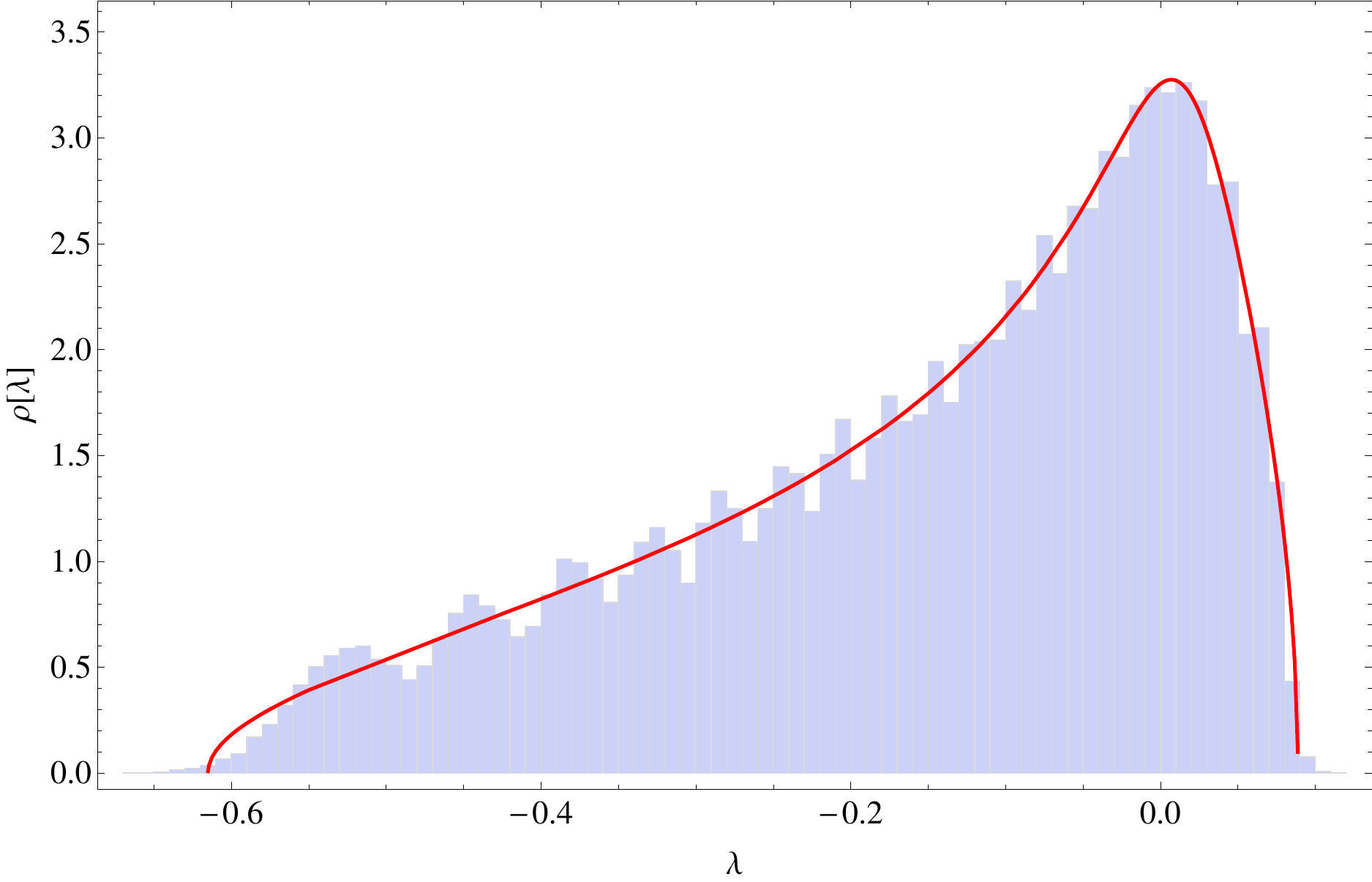}&\includegraphics[scale=0.28]{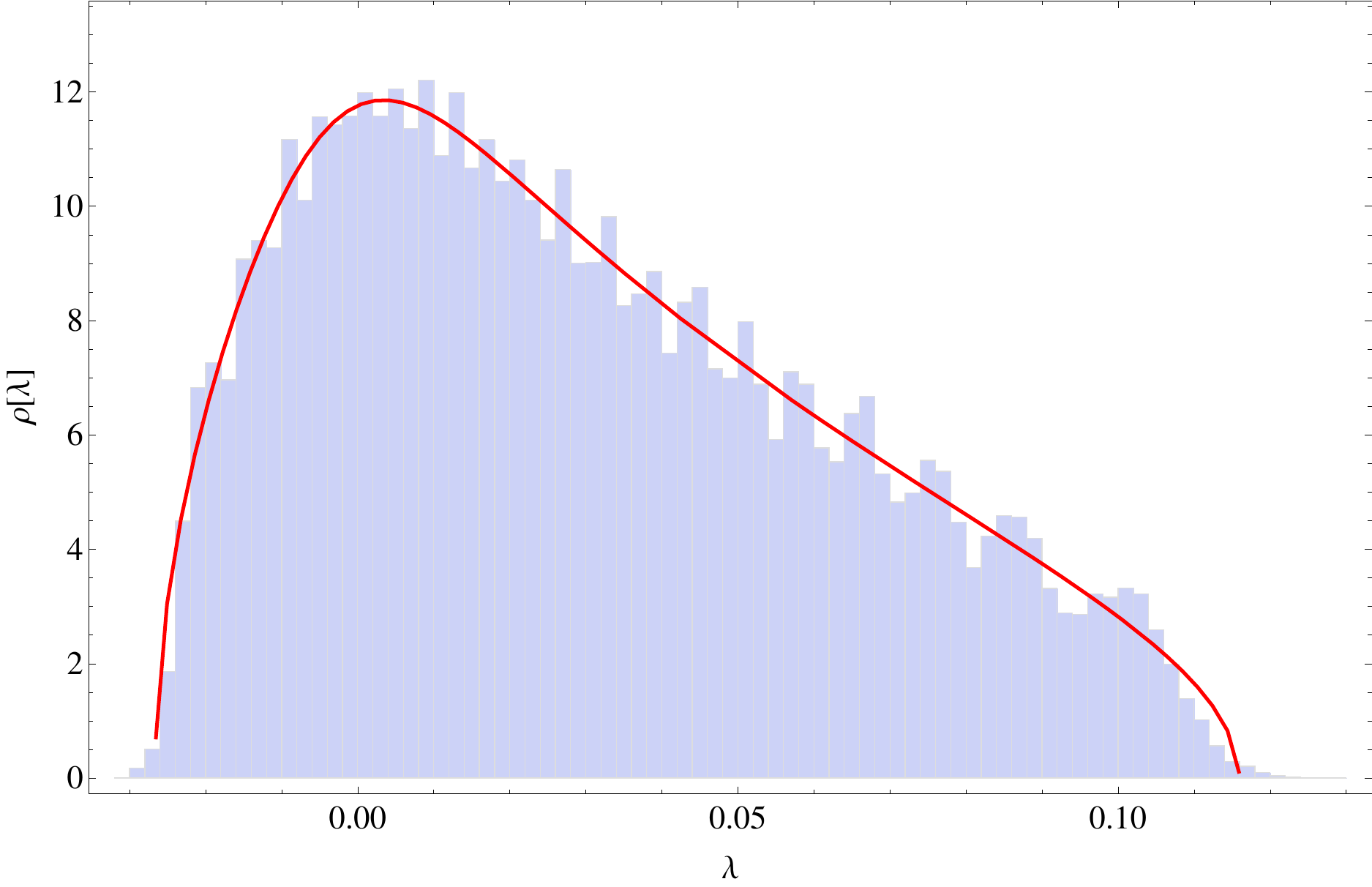}
	\end{tabular}
	\caption{Eigenvalue density: Theoretical results (red line) vs numerical results(blue bars) for the aymptotic case. (Left: $c=2/3$, $\eta=4$)(Right: $c=2/5$, $\eta=0.4$) }
	\label{comparation}
\end{figure}

 The main result of section \ref{section_Asymptotic} can be generalized to the case where the random matrices do not have the same weight in the difference. In this appendix we will show how to compute the AED for this asymmetric case, namely
\begin{equation}
Z=p\rho_1-q\rho_2.
\end{equation}
We can define a normalized matrix $\hat{Z}=Z/p$ and if $\eta:=q/p$, we can find the AED for
\begin{equation}
\hat{Z}=\rho_1-\eta\rho_2,
\end{equation}
and then go back to the original variables. We begin by calculating the $\mathcal{R}$ transforms for the AEDs of $\rho_1$  and  $-\eta\rho_2$ which we will call $\tilde{\density}$ and $\eta\tilde{\density}$ respectively. The $n$th cumulant is a homogeneous function of degree $n$, i.e., $k_n[\eta\tilde{\density}]=(-\eta)^nk_n[\tilde{\density}]$, thus the $\mathcal{R}$ transforms are:
\begin{eqnarray}
\mathcal{R}_{\tilde{\density}}(z)=\dfrac{1}{1-c z},\\
\mathcal{R}_{\eta\tilde{\density}}(z)=-\dfrac{\eta}{1+c\eta z},
\end{eqnarray}
where $c=N/M$ as in section \ref{section_Asymptotic}. The $\mathcal{R}$ transform for the AED of $\hat{Z}$ will be the sum of the $\mathcal{R}$ transforms, explicitly
\begin{equation}
\mathcal{R}(z)=\dfrac{1}{1-c z}+\left(-\dfrac{\eta}{1+c\eta z}\right)=\frac{(1-\eta )+ \eta (2c z)}{(1-c z)(1+\eta  c z)}.
\end{equation}
Following the same procedure as in section \ref{sectiondiff} we obtain an equation for the Cauchy transform $G(z)$
\begin{equation}\label{cauchyeta}
\frac{(1-\eta )+ \eta (2c G(z))}{(1-c G(z))(1+\eta  c G(z))}+\frac{1}{G(z)}=z.
\end{equation}
Equation \eqref{cauchyeta} is cubic on $G(z)$, analizing the possible roots with the same criteria of analyticity as in the symmetrical case we find the desired AED
\begin{small}
	\begin{equation}\label{newrho}
	\density(x,c,\eta)=\frac{1}{2\pi|x|c\eta }\left(w(x,c,\eta)-\frac{\left(1+\eta +\eta ^2\right)x^2 - (1+c) (\eta-1 )\eta x+(c-2)^2 \eta^2}{3 w(x,c,\eta)}\right),
	\end{equation}
\end{small}
with
\begin{small}
	\begin{equation}
	w(x,c,\eta)=\left(\sqrt{\frac{x^2 \eta ^2 }{4}\left(\left(x-x_{c,\eta}\right)^2-\alpha^2\right) \left(\left(x-x_{c,\eta}\right)^2-\beta^2\right)}-\frac{f(x,c,\eta)}{6\sqrt{3}}\right)^{\frac{1}{3}},
	\end{equation}
\end{small}
where
\begin{equation}
x_{c,\eta}=\frac{(1+c)(1-\eta)}{2}
\end{equation}
and
\begin{equation}
f(x,c,\eta)=((x-a)-\gamma)((x-a)-\bar{\gamma})((x-a)-\epsilon),
\end{equation}
where
\begin{equation}
a=\frac{(1+c) \eta  (1+\eta  (4+\eta ))}{(\eta -1) (2+\eta ) (1+2 \eta )}.
\end{equation}
The parameters $\alpha$, $\beta$, $\gamma$ and $\epsilon$ are functions of $c$ and $\eta$ and have cumbersome expressions which we will not write here.
\\
\\
The structure is very similar to the result obtained in \eqref{alternanive} with the difference that the limits $[x_-,x_+]$ are changed with $\eta$. Although it is a complicated expression for $\density$, the absolute maximum limit can be found in the case $\eta>1$. This limit is equal to $\frac{(1+c)(1-\eta)}{2}-\beta$. In Figs. \ref{compararion} and \ref{comparation}  we compare our result with numerical simulations, note that we have only considered here the case $c<2$.
\subsection{Appendix B}
In this appendix we derive average eigenvalue density \eqref{d2equation} for the case $N=2$. The first step is to note that for $N=2$, the two eigenvalues satisfy the condition $\lambda_1 = -\lambda_2$, hence there is only one independent eigenvalue. Using this constraint, Eqs. \eqref{distribucion} and  \eqref{diagdistintegral}, with $\gamma = 1 - |\lambda|$,  the PDF for the independent eigenvalue can be written as
\begin{equation}
\label{predist2}
\density(\lambda) = -\lambda \frac{d}{d\lambda} \left( \frac{\Gamma(2M)^2}{\Gamma(M)^{2N}} (1- |\lambda|)^{4M-3} W\left(\frac{|\lambda|}{(1 - |\lambda|)^2}\right)\right)\, .
\end{equation}
where
\begin{equation}
 W(\alpha) = \int_0^{1}\!dx\, \left (x(1-x) \right)^{M-1} \left(x(1-x) + \alpha\right)^{M-1}.
\end{equation}
Expanding the second factor in the integral in a binomial expansion, and using the Beta integral, $W(\alpha)$ can be expressed in terms of a hypergeometric function
\begin{eqnarray}
W(\alpha) & =  & \sum_{k=0}^{M-1} {M-1 \choose k} \frac{ \left( (2(M-1) - k)!)^2\right)}{(4(M-1) - 2 k +1)!}\alpha^k \\
                & = & \frac{ ((2M-1)! )^2 }{(4M-3)!} {}_2F_1 \left(\left.\begin{array}{c} 3/2\!-\!2M \ \ \ 1\!-\!M \\ 2(1\!-\! M) \end{array}\right |- 4\alpha  \right).
\end{eqnarray}
Using the standard hypergeometric identity
\begin{equation}
{}_2F_1\left(\left.\begin{array}{c} \!a \ \ \ b \\ c \end{array}\right |z  \right)=(1-z)^{-b}{}_2F_1\left(\left.\begin{array}{c} \!c\!-\!a \ \ \ b \\ c \end{array}\right | \frac{z}{z-1}  \right),
\end{equation}
$W(\alpha)$ can further  be simplified to
\begin{equation}
W(\alpha) =\frac{ ((2M-1)! )^2 }{(4M-3)!} (1 + 4 \alpha)^{M-1} {}_2F_1 \left(\left.\begin{array}{c} 1/2 \ \ \ 1\!-\!M \\ 2(1\!-\! M) \end{array}\right |\frac{ 4\alpha }{1 + 4 \alpha} \right) .
\end{equation}
Substituting into equation \eqref{predist2}, we finally obtain \eqref{d2equation} after simplification.

\end{document}